\documentclass[pre,aps,twocolumn,showpacs,floatfix,superscriptaddress]{revtex4-2}
\usepackage{graphicx}
\usepackage{bm}   
\usepackage{color,soul}
\usepackage{amssymb}
\usepackage{amsmath}
\usepackage{ulem}
\usepackage[dvipsnames]{xcolor}
\usepackage{subfigure}
\usepackage[dvipsnames]{xcolor}
\usepackage[toc,page]{appendix}
\usepackage{hyperref}
\hypersetup{colorlinks=true, citecolor=blue, urlcolor=blue, linkcolor=blue}

\newcommand{\e}{\normalfont\mbox{e}\,}

\begin{document}	
	\title{Intermittent cluster synchronization in a unidirectional ring of bursting neurons}
	\author{R. Sree Ardhanareeswaran}
	\affiliation{Department of Nonlinear Dynamics, Bharathidasan University, Tiruchirappalli 620024, Tamil Nadu, India}
	\author{S. Sudharsan}
	\email{hari.sudharsan32@gmail.com}
	\affiliation{Physics and Applied Mathematics Unit, Indian Statistical Institute, 203, B. T. Road, Kolkata - 700108, India}
	\author{M. Senthilvelan}
	\email{velan@cnld.bdu.ac.in (Corresponding Author)}
	\affiliation{Department of Nonlinear Dynamics, Bharathidasan University, Tiruchirappalli 620024, Tamil Nadu, India}
	\author{Dibakar Ghosh}
	\email{dibakar@isical.ac.in}
	\affiliation{Physics and Applied Mathematics Unit, Indian Statistical Institute, 203, B. T. Road, Kolkata - 700108, India}
	
	\begin{abstract}
		We report a new mechanism through which extreme events with a dragon king-like distribution emerge in a network of unidirectional ring of Hindmarsh-Rose bursting neurons interacting through chemical synapses. We establish and substantiate the fact that depending on the choice of initial conditions, the neurons are divided into different clusters. These clusters transit from a phase-locked state (anti-phase) to phase synchronized regime with increasing value of the coupling strength. Before attaining phase synchronization, there exists some regions of the coupling strength where these clusters are phase synchronized intermittently. During such intermittent phase synchronization, extreme events originate in the mean-field of the membrane potential. This mechanism, which we name as intermittent cluster synchronization, is proposed as the new precursor for the generation of emergent extreme events in this system. These results are also true for diffusive coupling (gap junctions). The distribution of the local maxima of the collective observable shows a long tailed non-Gaussian while the interevent interval follows the Weibull distribution. The goodness of fit are corroborated using probability-probability plot and quantile-quantile plot. This intermittent phase synchronization becomes rarer and rarer with the increase in the number of clusters of initial conditions.
	\end{abstract}
	\maketitle
	\section{Introduction}
	Uncertainties in the occurrences of exceptional natural events and the severe impacts caused by these rarities on the ecology and human life have constantly driven the interest of scientific communities to have a major focus on them. Such events which are rare and consequently disastrous are classified as extreme events (EE). Natural disasters like floods, tsunamis, earthquakes, and man-made crises such as power outages and nuclear accidents all fall under this classification \cite{extreme_book}. Measures to safeguard people and environment from these disasters are pivotal. So understanding the underlying complex mechanism becomes crucial in order to forecast such events and establish preventive strategies. To achieve this, depending on the problem at hand, appropriate dynamical models are investigated and the mechanisms underlying EE should be established. In this direction, from the past decade, researchers have started exploring the birth and origin of EE in a variety of physical, environmental, societal, biological, cosmological, optical \cite{optical,multistable} and experimental systems \cite{lab}.
	
	\par Despite the lack of an universally accepted definition, EE are classified by their abnormal characteristics. Statistical indicators like the significant height $H_s$ help to distinguish EE  from other regular events \cite{pot,pot1,pot2}. In a dynamic sense, EE are described as a sudden shift of a variable from a normal state to a high-amplitude state. A long-tail non-Gaussian probability distribution is typically seen when analyzing the amplitudes of such dynamic observables, indicating the presence of infrequent large-scale events \cite{review}. Sometimes dragon-king distribution has also been observed in the distribution patterns \cite{extreme_book}. Extreme events also called sometimes as ``Black swans" which arise as a continuous process in the tail of the power-law distribution. In contrast, ``Dragon-kings" is not a result of a continuous process and hence it does not fall in the tail of a power-law distribution, rather those events deviate from the power-law distribution and follows a separate non-Gaussian distribution.  Most small to medium-scale events (city populations, earthquake sizes, stock price fluctuations, etc.) follow a power law, except for sporadic huge events that deviate as outliers. On the other hand, it has been shown that the significant mechanisms behind these abrupt transitions in isolated systems are interior crisis \cite{ic1,ic2,ic3,ic4,ic5,crosscris}, Pomeau–Manneville (PM) intermittency \cite{ic1,route}, noise-induced intermittency \cite{noise}, sliding bifurcation \cite{iso1}, alternation in size of the chaotic attractor \cite{ic4} and strange nonchaotic attractor \cite{iso13,iso14}. Sudden expansions of chaotic attractors through interior crises lead to EE in complex systems \cite{icc1,icc2,icc3,icc4} as well. Current research also focus on exploring other potential causes of EE both in isolated and complex systems \cite{review,iso1,iso2,iso3,multistable,iso5,iso6,iso7,iso8,iso9,iso10,iso12,iso13,iso14,iso15,iso16,iso17,iso18}. Especially, in complex systems, a network of Fitzhugh-Nagumo neurons with global and small world coupling topology produces EE \cite{icc1,icc2}.  Similarly, EE also evolve in networks comprising of Josephson junctions \cite{iso7}, Li\'enard systems \cite{ic1}, Stuart-Landau oscillators \cite{stuart}, coupled chaotic maps \cite{glomap}, moving agents \cite{moving}, neuronal maps that are diffusively coupled locally and nonlocally \cite{roycoup}. These rare events manifest in complex coupled systems as a result of the excitability of proto events \cite{icc1}, instability of out-of-phase synchronization \cite{iso5}, the interplay of attractive and repulsive coupling \cite{moving}, riddled basing \cite{riddled}, on-off intermittency \cite{glomap,moving}, attractor bubbling \cite{iso3}, the interplay between degree distribution and repulsive coupling \cite{iso10}, and counter-rotating chaotic attractors \cite{stuart}. These are the very few mechanisms that have been determined so far.
	
	In the quest for a new precursor of EE, we come to know that several new mechanisms in a variety of coupled dynamical systems and several exemplary models existing in the literature are yet to be investigated. One such appealing model is the Hindmarsh-Rose (HR) neuron model \cite{hr}, a simple, nonlinear and a computationally feasible model, known to display almost all the dynamical properties of a neuron both as an isolated \cite{hr1,hr2} and as a coupled \cite{hr3,hrreview} system. Despite being a rich model, hitherto, analysis of EE on HR model is very few and countable. For instance, EE are found to occur in an isolated HR model due to a sudden hyperchaotic expansion of the attractor as a result of variation of the external current \cite{leoconf}. Further, in a memristive based HR (MHR) neuron model, superextreme and extreme spikings are found to originate at the point of interior crisis \cite{dineshsuper} and due to the presence of noise \cite{energyhr} respectively. Subsequently, in Refs.~\cite{iso5,route}, for a two coupled HR model, dragon kings are found to occur due to quasiperiodicity. For the same system, hyperchaos triggered by PM intermittency \cite{leotransition} is also a well explained precursor. Unidirectional and bidirectional coupling are also established to produce EE in a two coupled MHR neurons \cite{dineshtransition}. Very recently, in Ref.~\cite{energyhr}, extreme energy functions were observed in an integer and a fractional order isolated MHR neurons. In the HR neuron model variants, studies undertaken on EE so far is either in a single or in a two coupled system. Investigations pertaining to EE are yet to be carried out in higher dimensional coupled HR systems. Even simple coupling topologies like local, nonlocal and global have not been analyzed so far in the EE analysis of the HR neuron network. Studying the simplified network serves as an essential step to efficiently understand the complex network. 
    \par In this article, we carry out the dynamical analysis of EE and determine its precursor in a locally coupled network of HR neurons whose individual nodes are nonlinearly (through chemical synapse) connected with each other. Previously, it is known that extreme events appear near synchronization manifold or are associated with intermittent synchronization/desynchronization. The relating scenarios are (i) \textit{intermittent bubbling of the transverse variable,} where the transverse variable intermittently bubbles out from the invariant manifold \cite{iso3}, (ii) interruption of the synchronized behavior pertaining to slips in the phase difference leading to an \textit{imperfect phase synchronization}  \cite{icc1,icc2}, (iii) excursion of the trajectories in the transverse direction to the synchronization manifold \cite{icc3}, (iv) \textit{instability of out-of-phase synchronization}, where an instability arises in the desynchronization manifold causing the trajectories of two oscillators to intermittently synchronize \cite{iso5}, and (v) on-off intermittency manifesting as a result of  \textit{intermittent desynchronization of the oscillators}  \cite{glomap}. Along these lines, we unveil an important and a new mechanism that intermittent cluster synchronization among the neurons in the network produces EE. From our analysis, we find that for the chosen number of the clusters of initial conditions (ICs), correspondingly neurons in the network split into that many clusters. These clusters evolve desynchronously most of the time and occasionally synchronize for a short duration. During such an intermittent synchronization, EE are observed in the collective observable of the network. This scenario proposed in this article is a novel mechanism and is reported for the first time to the best of our knowledge. In general, systems with inherent symmetry is found to transit from antiphase to in-phase synchronization upon varying the coupling strength and symmetry in the initial conditions might lead to results that are not generalizable to more complex scenarios \cite{gb1,gb2,gb3,gb4,gb5}. In the present case, intermittently, in time for a particular value of the coupling strength, intermittent cluster synchronization occurs leading to extreme events in the mean field variable. This occurs irrespective of whether symmetry in the clusters of ICs is present in the system or not. Further, the rarity of the EE is found to increase with increase in the number of ICs.  Typically, intermittent synchrony is observed in brain neuron networks and the EE emerging in them are deeply related to epilepsy (like  epileptic seizures) \cite{epilepsy,epilepsy1,igf1} when a highly detrimental disorder happening in the human brain collapsing the individual. Present study and the mechanism found out of this work will serve as an important milestone in the understanding of dynamical mechanisms behind epileptic seizures.
	\par We present our work as follows. In Sec.~II, we describe the model that we take for investigation, discuss the emergence of EE in particular network topology, detail the statistics followed by them and discuss about the spiking to bursting transition. In Sec.~III, we demonstrate the mechanism behind the events. The Robustness of EE to the ICs is studied in Sec.~IV. Finally in Sec.~V, we summarize and conclude the findings.	
	\section{Locally Coupled Network}
	The HR neuronal model \cite{hr} is well known for the chaotic and different types of bursting behaviour \cite{hr3}. Despite being a non-physiological model (due to the cubic nonlinearity), HR model serves to be a prototypical model for bursting and helping us to qualitatively understand the nature of neuron's dynamics in an easier way when compared to the other physiological models.  The mathematical form of the HR model is expressed as
	\begin{equation}
		\begin{split}
			\dot{x}&=y+bx^2-ax^3-z+I, \\
			\dot{y}&=c-dx^2-y,\\
			\dot{z}&=r[s(x-x_r)-z],\\
		\end{split}  
		\label{eqnsingle}
	\end{equation}
	where the variable $x$ represents the membrane potential, $y$ and $z$ represent the transport of ions in the fast and slow channel respectively. The variable $y$ describes the dynamics of fast current, while the variable $z$ characterizes the dynamics of the slow current. The speed of the slow variable $z$ is controlled by the parameter $r$. The parameter $I$ gives the external current that enters the neuron. $x_r$ is the parameter that controls the delaying and advancing activation of the slow current in the neuron. Other parameters, namely $a,~b,~c,~d,~\text{and},~S$ are positive coefficients. 
	
	\begin{figure}[h!]
		\centering
		\includegraphics[width=0.7\linewidth]{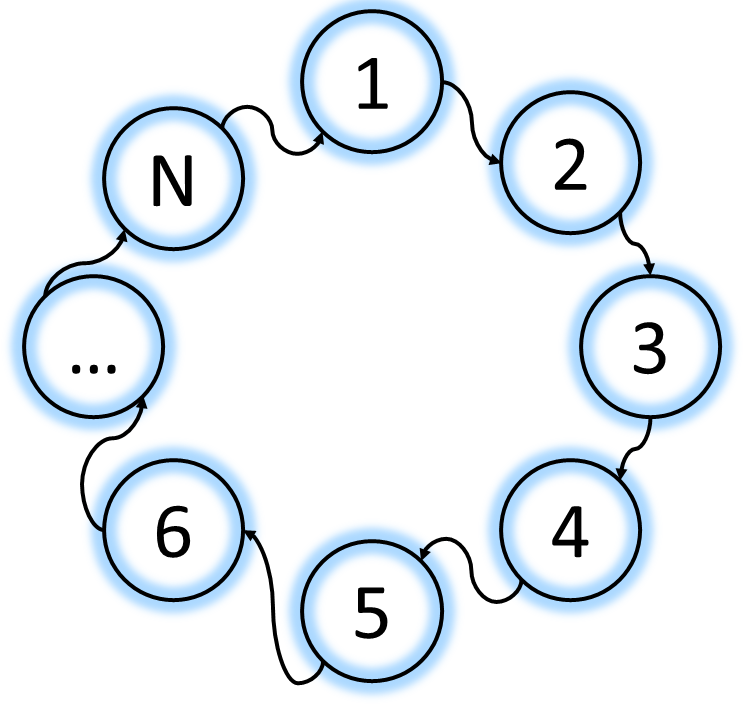}
		\caption{Schematic diagram of the $N$ unidirectionally coupled systems. Each numbering corresponds to a node, each evolving according to Eq.~(\ref{eqnsingle}).}
		\label{fig:1}
	\end{figure}
	
	 Now let us consider $N$ identical copies of HR neurons \cite{hr3,hrnet} which are coupled in a unidirectional ring topology as shown in the schematic Fig.~\ref{fig:1}. Modifying Eq.~(\ref{eqnsingle}) for this schematic, we get
	\begin{equation}
		\begin{split}
			\dot{x}_i&=y_i+bx_i^2-ax_i^3-z_i+I+k(v_s-x_i)\Gamma(x_{i+1}), \\
			\dot{y}_i&=c-dx_i^2-y_i,\\
			\dot{z}_i&=r[s(x_i-x_r)-z_i],\\
		\end{split}  
		\label{eqncoupled}
	\end{equation}
	where $i$ is the number of neurons in the system with the boundary condition ($x_{N+1}=x_1$) and $N$ is the total number of neurons. In our analysis, we take $N=100$. With this boundary condition, the chain of locally coupled oscillators looks like a ring, as visible from Fig.~\ref{fig:1}. $i^{th}$ oscillator is coupled to the $(i+1)^{th}$ oscillator nonlinearly through the sigmoidal function ($\Gamma(x)$): a type of coupling function which is commonly used to represent the chemical synaptic interaction \cite{hrreview}. The mathematical form of the sigmoidal function \cite{hr3} is given by  
	\begin{eqnarray}
		\Gamma(x_j)=\frac{1}{1+\e^{-\lambda(x_j-\theta)}}. 
		\label{sigmoidal}
	\end{eqnarray}
	The parameters $\lambda$ and $\theta$ in Eq.~(\ref{sigmoidal}) correspond to the slope of the function and the synaptic threshold respectively. Throughout our study, we fix the values of the system parameters at $a=1, b=3, c=1, d=5, x_r=-1.6, r=0.01, s=5, v_s=2$, $\lambda=10$, $\theta=-0.25$. $I$ is the current given to the neurons and $k$, the coupling strength between the neurons, is taken as the bifurcation parameter. For the chosen set of parameters, the uncoupled neuron exhibits three periodic dynamics.
	
	\par 
	
	Here, we investigate two cases, namely inhibitory coupling where $k$ is negative and excitatory coupling where $k$ is positive. Subsequently, in the following, we will discuss EE resembling dragon kings appearing in the system (\ref{eqncoupled}). Equation (\ref{eqncoupled}) is integrated numerically using the fifth-order Runge-Kutta-Fehlberg integration scheme with a step length $\Delta t=0.01$ \cite{github}. We have considered the HR neuron model since this model gives a variety of dynamics by changing the parameter I. Since, the understanding of collective behaviours in complex neuronal networks is a challenging task, so for simplicity, we have chosen ring type of HR neurons where the coupling is unidirectional. Since, the ring type of network topology is symmetric, we can consider two distinct ICs, odd and even. For our numerical analysis, a set of initial conditions is used to initialize the state variables. Among the $N=100$ neurons, odd neurons ($i=1, 3, ..., 99$) are set to the initial conditions (IC's): $x_{odd}=1, y_{odd}=0.5, z_{odd}=1$, and the even neurons ($i=2, 4, ..., 100$) are set to the IC's: $x_{even}=0.5,y_{even}=1,z_{even}=0.5$. We consider the mean of the $x$ variable $ X_m = \frac{1}{N}\sum_{i=1}^{N} x_i$ as the observable. Similarly $ Y_m = \frac{1}{N}\sum_{i=1}^{N} y_i$ and $ Z_m = \frac{1}{N}\sum_{i=1}^{N} z_i$.
	
	\subsection{Emergence of extreme events}
	\par We start by using the peak over threshold method \cite{review} and classify the events that are larger than a predefined threshold level ($H_s$) as EE. The threshold level is estimated using
	\begin{equation}
		H_s=\mu+n\sigma, \hspace{0.5cm} n \in \mathbb{R}~\backslash~\{0\}~\text{and}~n>1,
	\end{equation} 
	where $\mu$ is the mean value and $\sigma$ is the standard deviation of the collective observable $X_m$. In our case, we chose $n=6$.
	
	\par
\begin{figure}[h!]
		\centering
		\includegraphics[width=1\linewidth]{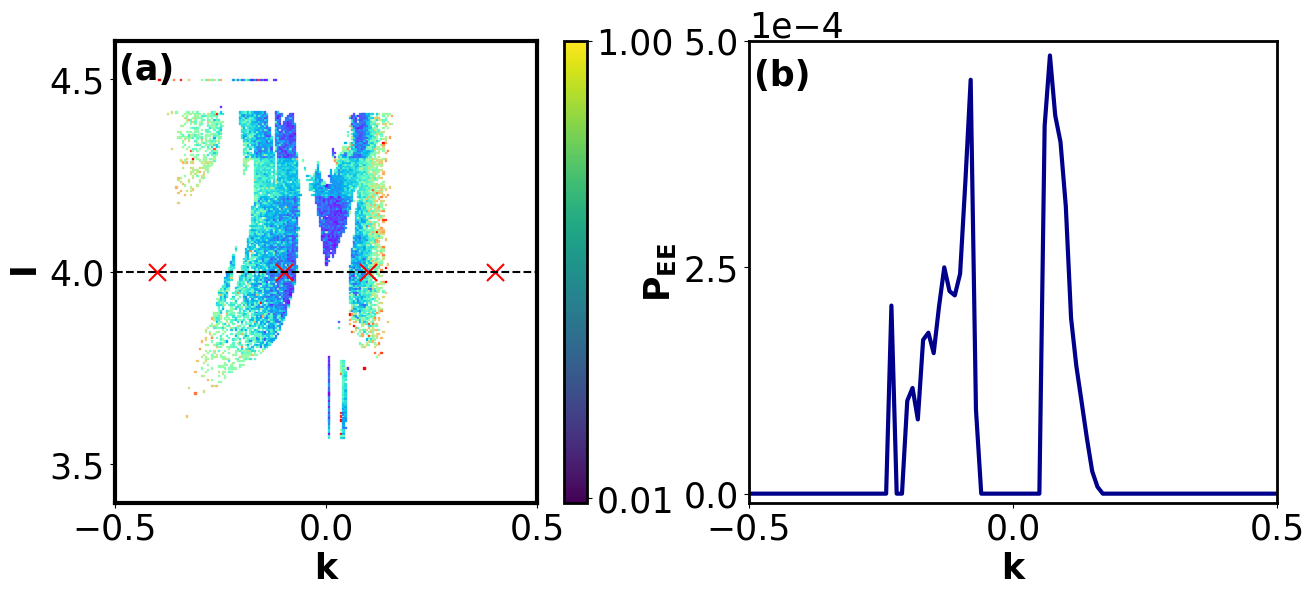}
		\caption{The probability of occurrence of EE in system (\ref{eqncoupled}) (a) in two-parameter ($k-I$) space and (b) in one parameter $k$ space. In (a), the coloured zones blue to lime green (black to light grey) represents probability of EE in ascending probability.}
		\label{fig:2}
	\end{figure}

 We perform numerical integration of system (\ref{eqncoupled}) up to $5\times10^{7}$ iterations of which $2\times10^{6}$ iterations were left out as transients and then the we plot the probability of the EE ($P_{EE}$) in the $(k-I)$ parameter space in Fig.~\ref{fig:2}(a). The calculation is carried out using the relation  
	\begin{eqnarray}
		P_{EE}=\frac{\text{Number of points crossing the threshold}}{\text{Total number of points}}.
	\end{eqnarray}  

	\begin{figure}[!ht]
	\centering
	\includegraphics[width=0.85\linewidth]{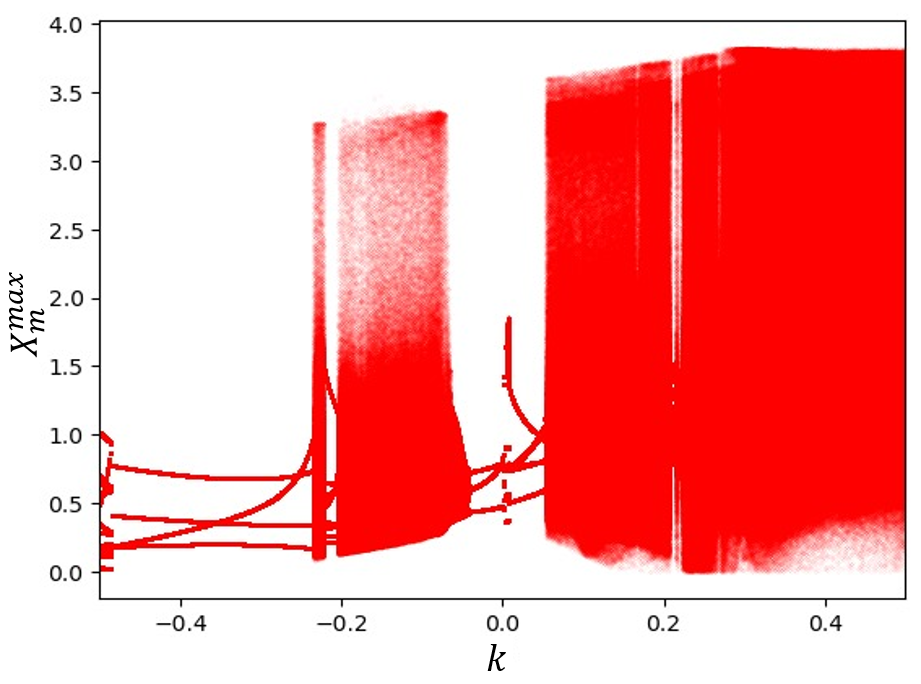}
	\caption{Bifurcation dynamics of the collective observable with respect to the coupling strength $k$ at $I=4.0$.}
	\label{collective}
   \end{figure}
	
	The colour bar in Fig.~\ref{fig:2}{a} displays the $P_{EE}$, where lime green (grey) represents the highest probability and dark blue (black) represents the lowest probability. In the white regions, no EE occur. Further observation of Fig.~\ref{fig:2}{a} reveals that more regions from the inhibitory coupling side are prone to produce EE compared to the areas from the excitatory coupling direction. Between the zones $I=3.6$ and $3.85$, we can see a short straight region in the coupling strength $k=0.04$ and $0.05$ where EE occur more than other regions.  We investigate these EE's birth and other statistical properties by discerning the region covered by the constant ordinate $I=4$. When we look at the region $I=4$ in Fig.~\ref{fig:2}(a), EE appear almost symmetrically in both directions of the coupling strength. To confirm this, we plot in Fig.~\ref{fig:2}(b), $P_{EE}$ with respect to the coupling strength $k$ at $I=4$. Increasing $k$, we can observe EE appearing in the system in both inhibitory and excitatory coupling strengths $k$. Figure~\ref{fig:2}(b) is also obtained for $5\times10^{7}$ iterations after leaving out transients of $2\times10^{6}$ iterations.
	From Fig.~\ref{fig:2}(b), we can see that EE occur for inhibitory coupling from $k=-0.23$ to $k=-0.06$ and from $k=0.06$ to $k=0.17$ for excitatory coupling.
	
	In Fig.~\ref{collective}, the bifurcation plot of the collective observable $X_m$ with respect to $k$ at $I=4$ is presented. We can observe that the system transits from periodic to chaotic nature upon increasing the coupling strength. This is true for both inhibitory as well as excitatory couplings. Upon comparing Fig.~\ref{fig:2}(b) and Fig. \ref{collective}, we can observe that while increasing the value of the coupling strength $k$ from $-0.5$, whenever EE arises in the system (i.e., non-zero $P_{EE}$ in Fig.~\ref{fig:2}(b)), the collective dynamics exhibits chaos (Fig.~\ref{collective}) and occasional in-phase emerges between two clusters. For the periodic regime in the bifurcation diagram in Fig. \ref{collective}, the probability of EE is zero due to the antiphase synchronization between the two clusters. The similar phenomena is also true for diffusive coupling case as well, which is discussed in the Appendix. The detailed mechanism of the emergence of EE is discussed in Sec.~\ref{birth}.

\begin{figure}
		\centering
		\includegraphics[width=1\linewidth]{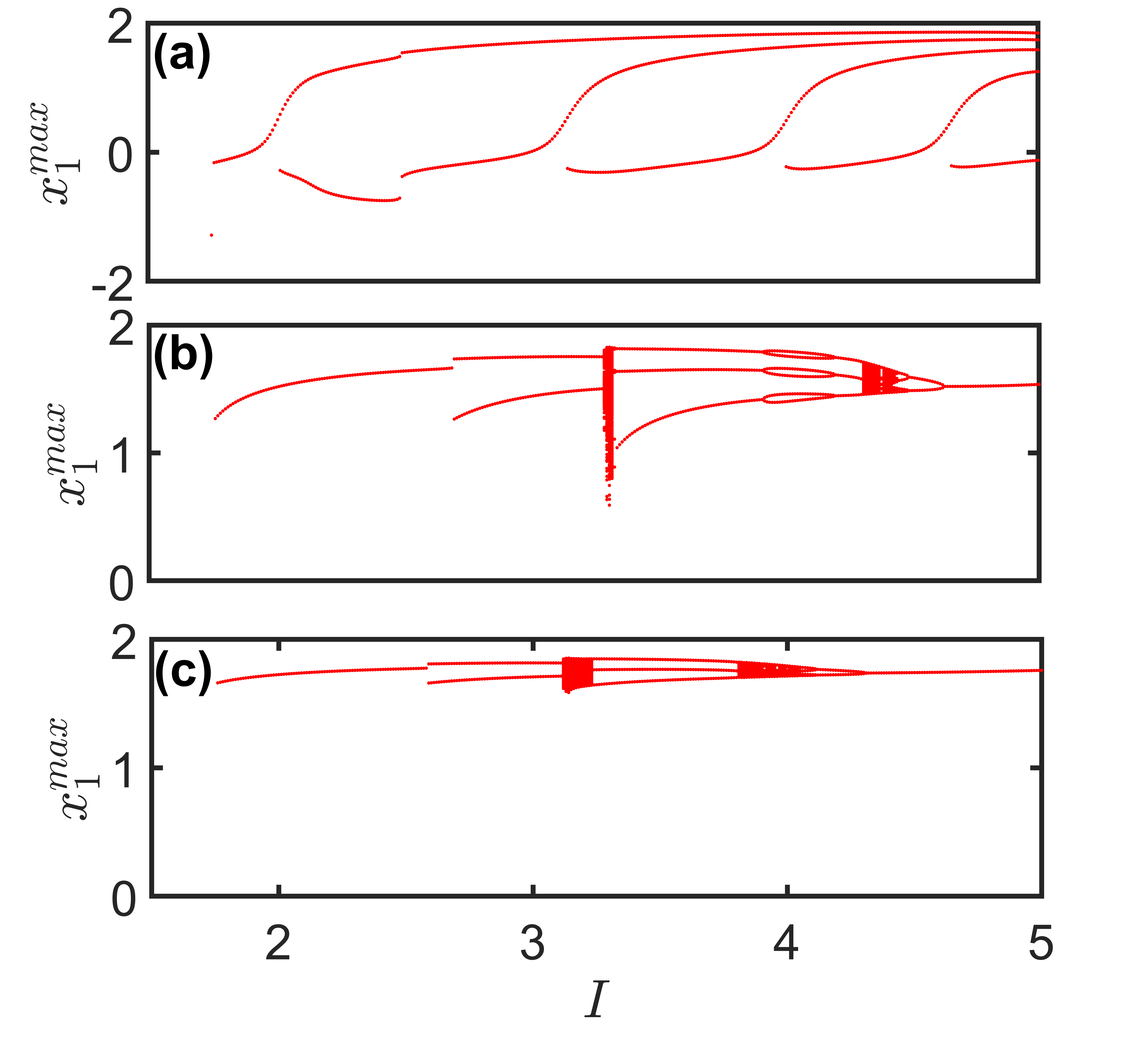}
		\caption{Bifurcation diagrams of the node-$1$ with respect to $I$ at the coupling strength (a) $k=-0.4$, (b) $k=-0.17$, (c) $k=0.4$.}
		\label{ibifur}
	\end{figure}

    Now, to see the local dynamics of the neurons, we plot the bifurcation diagram of individual neuron (say, 1) of system (\ref{eqncoupled}) respectively for three different values of the coupling strength $k=-0.4$, $k=-0.17$ and $k=0.4$ in  Figs.~\ref{ibifur}(a-c). We can observe the transition from periodic to the chaotic states upon increasing the coupling strength $k$. At $k=-0.4$, the dynamics is completely periodic whereas, at $k=-0.17$ and $k=0.4$, the periodic states are interspersed by chaotic states through period doubling route. Individual neurons may exhibit periodic or chaotic dynamics, but for appropriate coupling strengths, intermittent cluster synchronizations occurs among the clusters of neurons.
	
	\subsection{Statistical showcase}
	In the previous subsection, we discussed the genesis of EE in the system (\ref{eqncoupled}). Now, we discuss the statistics followed by them. In this section, all the analyses are carried out for about $10^9$ iterations, leaving out $10^5$ iterations as transients.
	
	\begin{figure}[h!]
		\centering
		\includegraphics[width=0.95\linewidth]{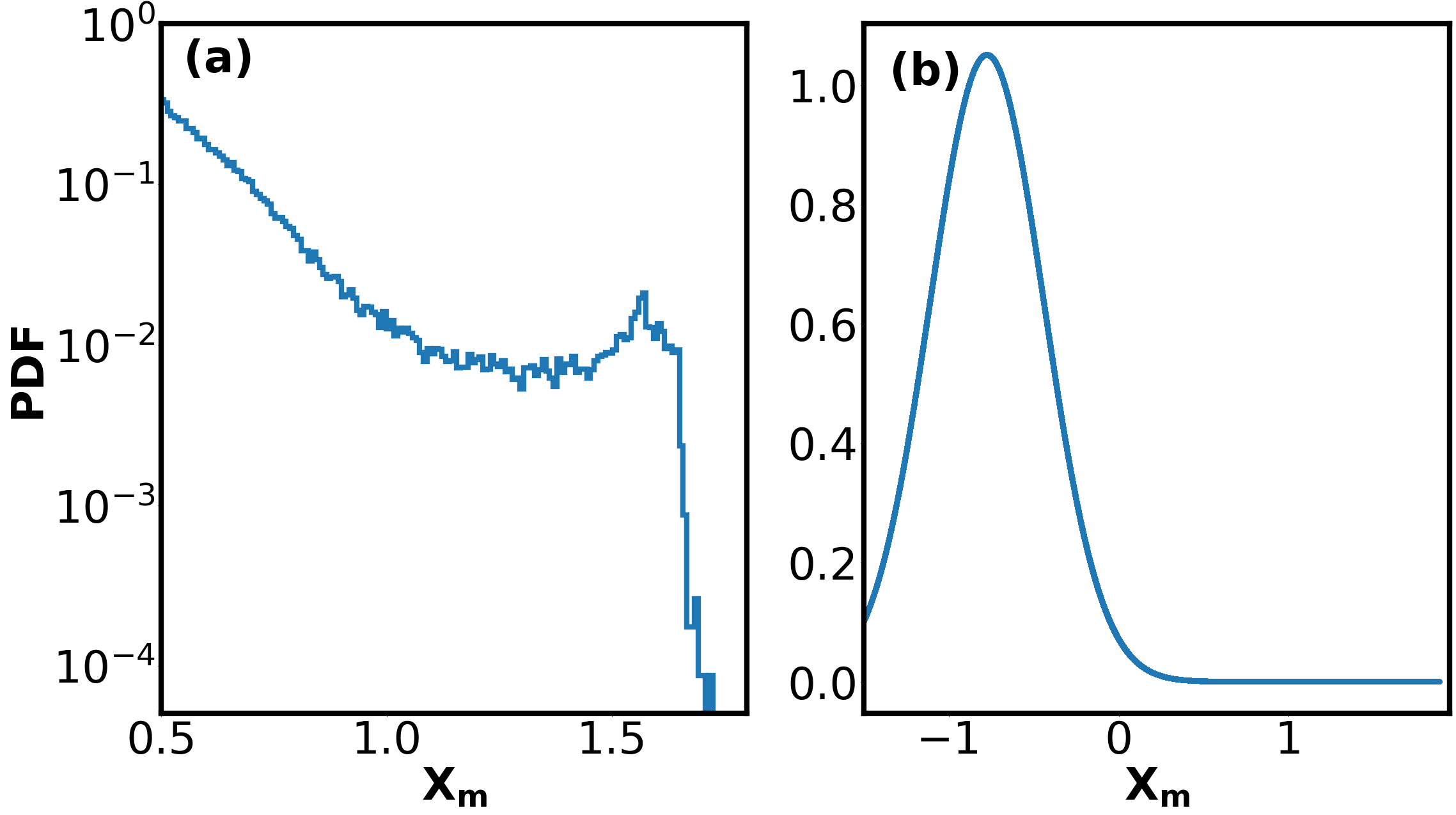}
		\includegraphics[width=1\linewidth]{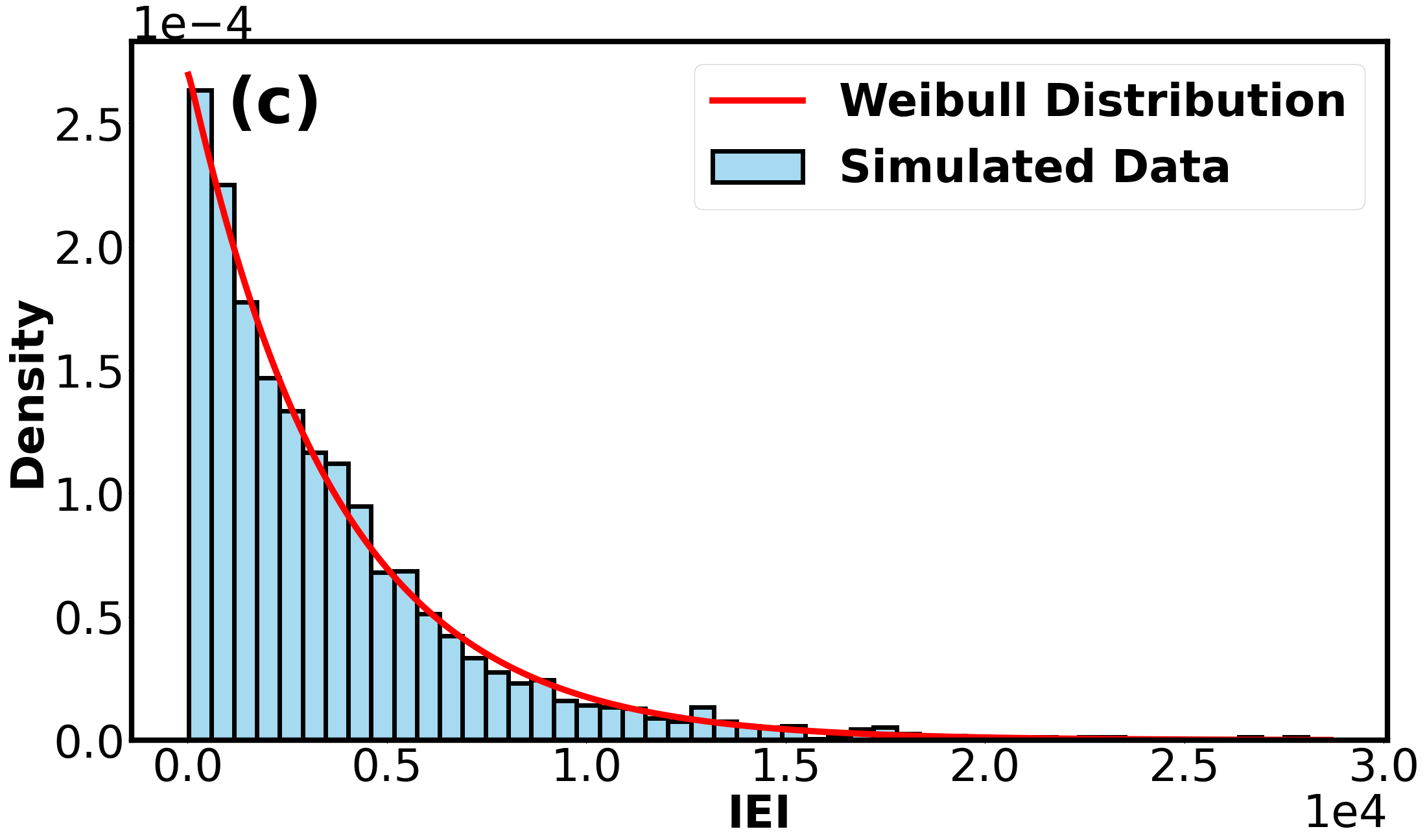}
		\includegraphics[width=0.95\linewidth]{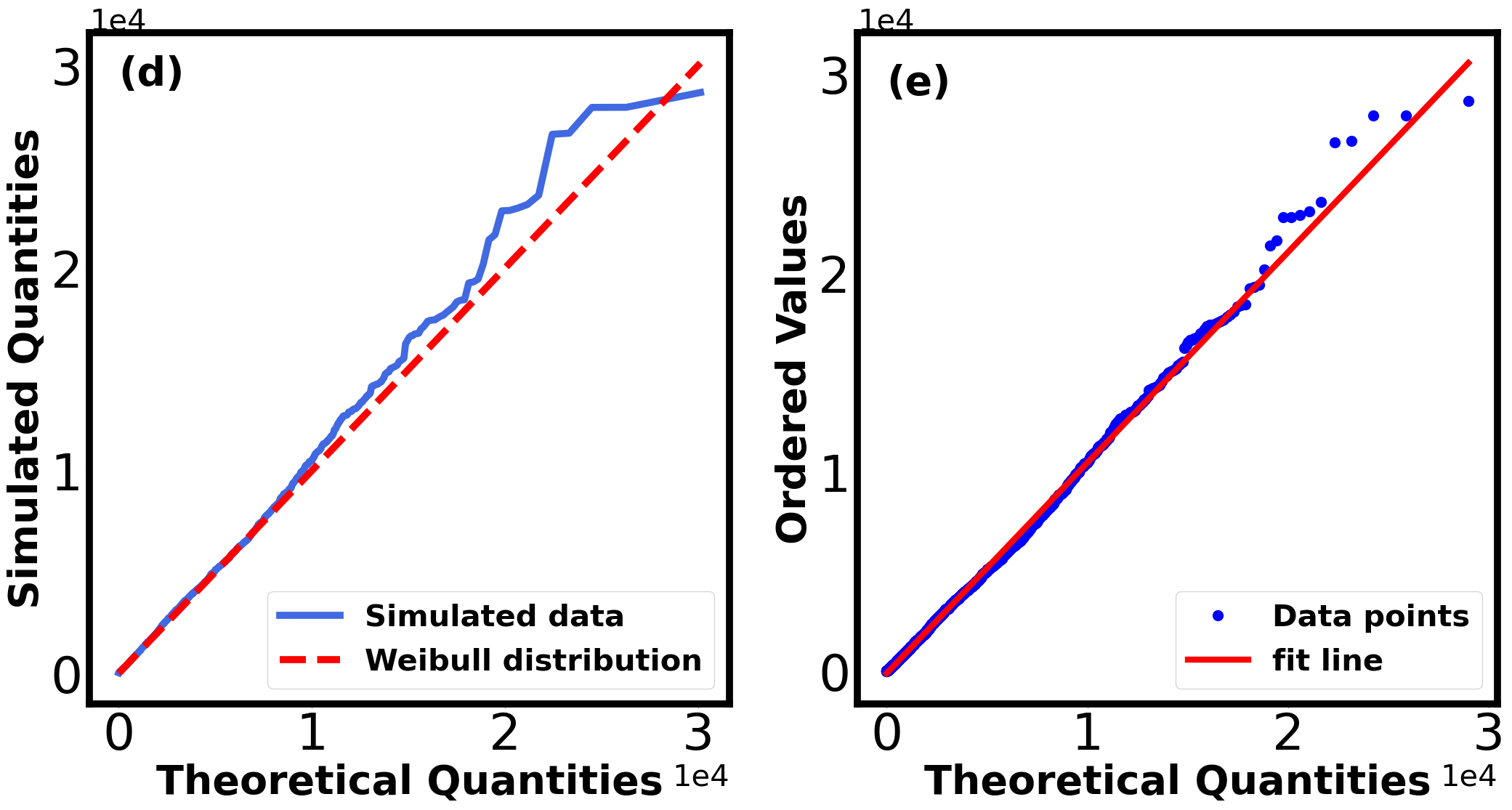}
		\caption{Distribution and statistics of EE of system (\ref{eqncoupled}) for the value of $k=-0.17$. Specially, (a) is the PDF plot showing the dragon king distribution, (b) is normal distribution of the entire data where long tail is found, Subplot (c) is the IEI histogram fitted with Weibull distribution. (d) and (e) are the PP and QQ plots, respectively. }
		\label{fig:12}
	\end{figure}
	
	In Fig.~\ref{fig:12}(a), we plot the probability distribution function of peaks by collecting the local maxima. From the plot, we can observe a slight hump at the back. From this, we can infer that the distribution of local maxima follows a dragon king-like distribution. Hence EE occurring in the system (\ref{eqncoupled}) are dragon king-like EE. Further, in Fig.~\ref{fig:12}(b), we plot the non-Gaussian distribution of the peaks where the distribution can be found to have a very long tail. This tail represents the EE and the probability of these tail events is near to zero but still non-zero. This represents the rarity of these events. Despite being rare, the consequences of these events are catastrophic. Dragon-kings basically exhibit clustering property. That is, very high probability for its occurrences beyond the threshold. This property makes them to significantly deviate from the power-law or any other normal distribution. In a system like network of neurons, where EE represent epileptic seizures, such a rare and high magnitude tail event is lethal. Next, we plot the inter-event interval (IEI) in Fig.~\ref{fig:12}(c). We find that Weibull Distribution fits very well with the IEI. These fitted values were acquired using the python package ``scipy.stats" where we acquired the values of the shape and scale parameters respectively as $1.00286$ and $3522.28717$.  We used the Freedman-Diaconis rule to set the number of bins to our simulated data sets \cite{glomap}. Figures~\ref{fig:12}(d) and \ref{fig:12}(e) represent the probability-probability (PP) and quantile-quantile (QQ) plots which decide the quality and correlation of the fit with the actual data sets. In Fig.~\ref{fig:12}(d), blue solid line is the cumulative distribution function (CDF) of the simulated data sets and the red dashed line is the CDF of the fit. From this, we can observe most of the times the fitted and simulated data sets correlate with each other which shows the quality of the fit. Figure~\ref{fig:12}(e) is the QQ plot. In this, the simulated data sets are considered as the cut points which divide the range of the continuous interval with equal probabilities. We used "PROBPLOT FUNCTION IN PYTHON" to plot the QQ plot. This function essentially estimates a probability plot for the given data with respect to the quantiles of the utilized distribution.

	\subsection{Spiking to Bursting Transition}
	\begin{figure*}[!ht]
		\centering
		\includegraphics[width=1.0\linewidth]{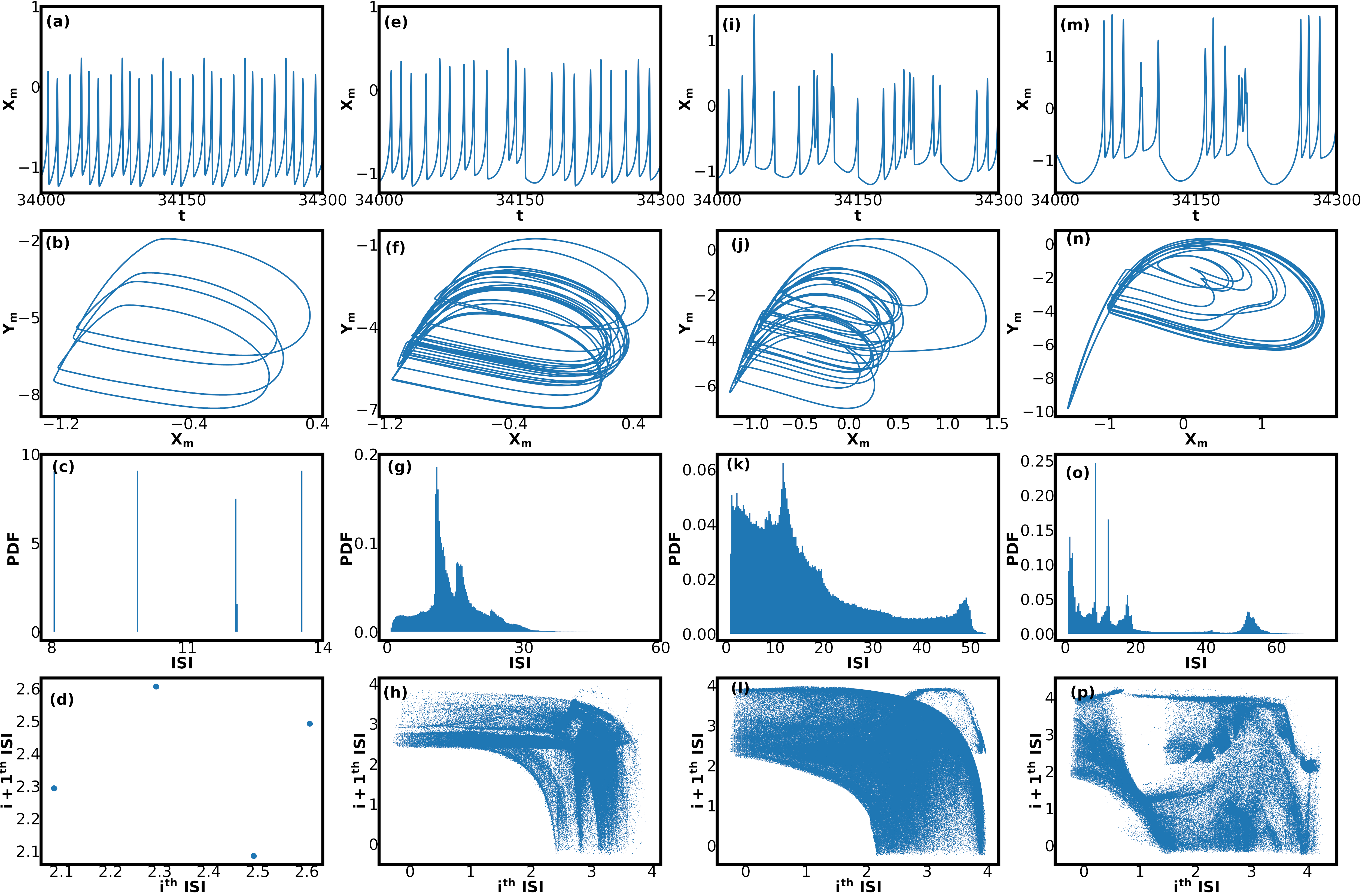}
		\caption{Time series (row 1), phase portrait (row 2), Interspike interval (ISI) distribution (row 3) and joint ISI histograms (row 4) of the collective observable of the system (\ref{eqncoupled}) for the coupling strengths $k=-0.4$ (column 1), $k=-0.1$ (column 2), $k=0.1$ (column 3) and $k=0.4$ (column 4).}
		\label{fig26}
	\end{figure*}
	
	Upon examining the dynamics of the system (\ref{eqncoupled}), we can understand that on changing the coupling strength $k$, dynamics of the collective observable $X_m$ make a transition from the spiking to bursting. To confirm this, we plot in Fig.~\ref{fig26}, the time series (Row 1), phase portrait (Row 2), interspike interval (ISI) distribution (Row 3) and joint ISI histograms (Row 4). This analysis is carried out and portrayed for four different values of the coupling strength, in particular $k=-0.4$ (Column 1), $k=-0.1$ (Column 2), $k=0.1$ (Column 3) and $k=0.4$ (Column 4). These four values of $k$ are considered at $I=4.0$ from the two parameter probability plot (see red cross marks in Fig.~\ref{fig:2}(a)) Initially, for the coupling strength $k=-0.4$, from the time series (Fig.~\ref{fig26}(a)) and phase portrait (Fig.~\ref{fig26}(b)), we can understand that the system exhibits a four periodic spiking behaviour. Correspondingly, in the PDF plot (Fig.~\ref{fig26}(c)), we can observe only 4 lines which confirm it is four-periodic. This is further confirmed by four dots in the joint ISI histogram (Fig.~\ref{fig26}(d)). Next for $k=-0.1$, observing the time series (Fig.~\ref{fig26}(e)) and phase portrait (Fig.~\ref{fig26}(f)), we can infer that at this zone the system exhibits chaotic spiking behaviour. This coupling strength lies in the EE emerging zone. The PDF plot (Fig.~\ref{fig26}(g)) resembles a Gaussian-type distribution which indicates the spiking is tonic \cite{spikeburst}. This is also confirmed by the joint ISI histogram (Fig.~\ref{fig26}(h)) where the joint ISI histogram forms an inverted L shape near the top right corner. Further, as we move down to the coupling strength $k=0.1$, the time series (Fig.~\ref{fig26}(i)) shows combined chaotic spiking and bursting. Phase portrait (Fig.~\ref{fig26}(j)) also confirms the mixture of these two dynamics. In particular, a small tail in the phase portrait ensures the presence of chaotic bursting as well. This is confirmed by ISI distribution (Fig.~\ref{fig26}(k)) which looks similar to the bursting form even though there is a small Gaussian peak which shows that the dynamics have alternative chaotic bursting and spiking. In the joint ISI histogram (Fig.~\ref{fig26}(l)), we can observe the same pattern as in the previous coupling strength. We can notice that in the EE zones, the patterns are similar. Finally, for $k=0.4$, we can observe purely bursting property (Fig.~\ref{fig26}(m)) which is slightly chaotic and the phase portrait (Fig.~\ref{fig26}(n)) confirms this as bursting. ISI PDF (Fig.~\ref{fig26}(o)) and joint ISI histogram (Fig.~\ref{fig26}(p)) also confirm that the dynamics is bursting type. In the ISI PDF plot (Fig. \ref{fig26}(o)), we can notice there are large probability for spikes having intervals between 0 and 20, followed by which there is a brief non-zero probability between 40 and 60 representing that there are no spikes present for that ISI. Then, there emerges a very tiny non-zero probability for the emergence of spikes intervaling between 40 and 60. Here, the non-zero probabilities for spike intervals ranging from 0 to 20 corresponds to the ISI within bursts. Whereas the very tiny non-zero probabilities for the spike intervals between 40 and 60 corresponds to the interburst ISI. Correspondingly, when we observe the joint ISI histogram (Fig.~\ref{fig26}(p)), we can see the intervals spread all over the regions confirming the bursting property.
	
	\par In the above, we have demonstrated that the collective variable $X_m$ undergoes a transition from spiking to bursting for increasing coupling strength $k$. Specifically, EE occur in two regions, namely regions of chaotic spiking (column 2 in Fig.~\ref{fig26}) and regions of alternate spiking and bursting (column 3 in Fig.~\ref{fig26}). Now we intend to investigate the mechanism by which system (\ref{eqncoupled}) gives birth to EE. Before proceeding further, it is important to note that due to the choice of the initial conditions, odd neurons with one set of IC's get synchronized separately and form one cluster and the even neurons with another set of IC's get synchronized separately and form another cluster. At any point in time, the formation of two clusters is apparent. Due to the property of synchronous evolution of neurons inside a single cluster, we choose one neuron from each cluster and study the nature of time series from the two neurons alone. This will allow us to understand the dynamics between the clusters easily. So, we choose neuron $1$ from the odd cluster and neuron $2$ from the even cluster and name them henceforth as $x_o$ and $x_e$ respectively. 
	
	\section{Birth of Extreme Events: Local Mechanism}
    \label{birth}
	To understand the mechanism, from the probability plot in Fig.~\ref{fig:2}(b), we consider three cases of the coupling strength, namely (i) $k=-0.4$, (ii) $k=-0.17$, and (iii) $k=0.4$ and recognize the mechanism by discerning the dynamics (microscopic) between EE and non-EE regions. By doing so we will be able to know, what phenomena at the micro level is causing EE at the macro level.
	\par
\begin{figure}
		\centering
		\includegraphics[width=1\linewidth]{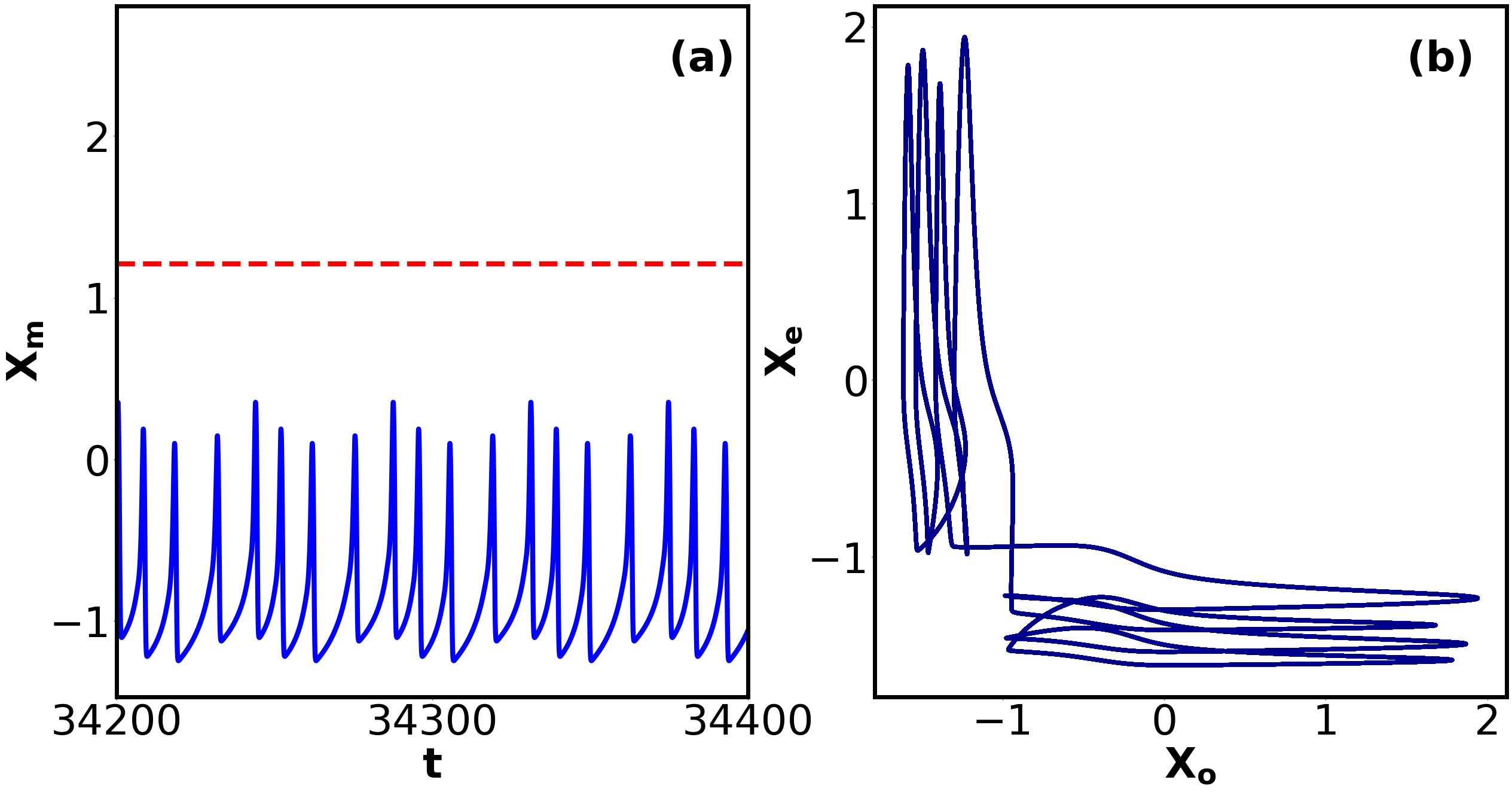}
		\includegraphics[width=1.05\linewidth]{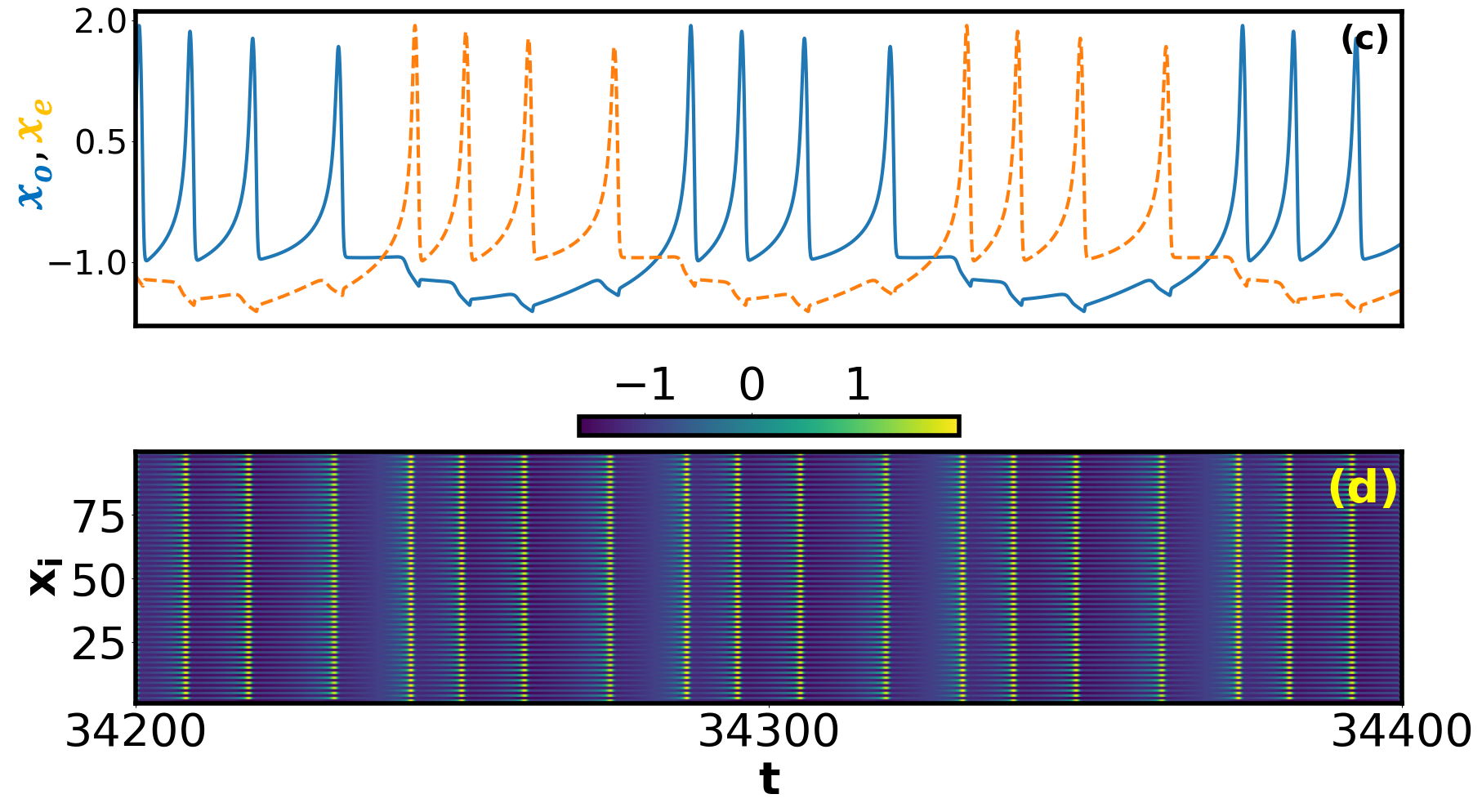}
		\caption{Macroscopic and microscopic dynamics of system (\ref{eqncoupled}) for $k=-0.4$. In particular, (a) $\&$ (b) are respectively the collective time series and synchronization error plot in the ($x_o-x_e$) plane. (c) represents the overlap of individual time series of $x_o$ (blue solid) and $x_e$ (red dashed). (d) is the spatio-temporal plot.} 
		\label{fig:k=-04}
	\end{figure}

 \textbf{(i) $k=-0.4$:} To understand in detail about the collective and individual dynamics, we plot in Fig.~\ref{fig:k=-04} the dynamics of the system (\ref{eqncoupled}) at $k=-0.4$. Specifically, Fig.~\ref{fig:k=-04}(a) is the time series of the collective observable $X_m$, Fig.~\ref{fig:k=-04}(b) represents the synchronization error plot of odd and even neurons, Fig.~\ref{fig:k=-04}(c) represents the time series of the individual neuron $x_o$ (blue solid line) overlapped with the time series of $x_e$ (orange dotted lines) and  Fig.~\ref{fig:k=-04}(d) is the corresponding spatio-temporal plot.  Time series in Fig.~\ref{fig:k=-04}(a), has no trajectories in the collective observable $X_m$ crossing the threshold $H_s= m+6\sigma=1.21$ line (dotted red horizontal line). If we carefully look at the time series plot of the individual clusters $x_o$ and $x_e$ (Fig.~\ref{fig:k=-04}(c)), we can notice that cluster $x_o$ bursts exactly in antiphase with the cluster $x_e$. Here odd neurons burst when the even neurons are at rest and vice versa. This is apparent from Fig.~\ref{fig:k=-04}(c). Further, both the clusters exhibit bursting with four spikes within each burst. This along with the antiphase evolution is the reason why the collective observable $X_m$ exhibits a four periodic spiking trajectory. Figure \ref{fig:k=-04}(b) gives us the overview of the dynamics in the ($x_{o}-x_{e}$) phase plane. We can see that no trajectories are in the transverse direction which is nothing but the synchronization manifold. This also confirms the antiphase synchronization occurring between them. Because of this, no trajectories of the collective observable $X_m$ cross the threshold. So, the probability of getting EE is zero. Next from the spatio-temporal plot in Fig.~\ref{fig:k=-04}(d), we can spot the oscillations of each and every neuron. Here we can easily notice that odd neurons and even neurons are not synchronously evolving in time. At any given point in time, we can notice that the plot has alternate green dots and violet zones, that is a discontinuity in those green dots in the vertical direction. The green and violet regions are excitation and  resting states, respectively. Alternate green and violet represents the spiking and resting states of alternate oscillators. This confirms the lack of synchronization (antiphase) between the odd and even neurons.
	
	\begin{figure}
		\centering
		\includegraphics[width=1\linewidth]{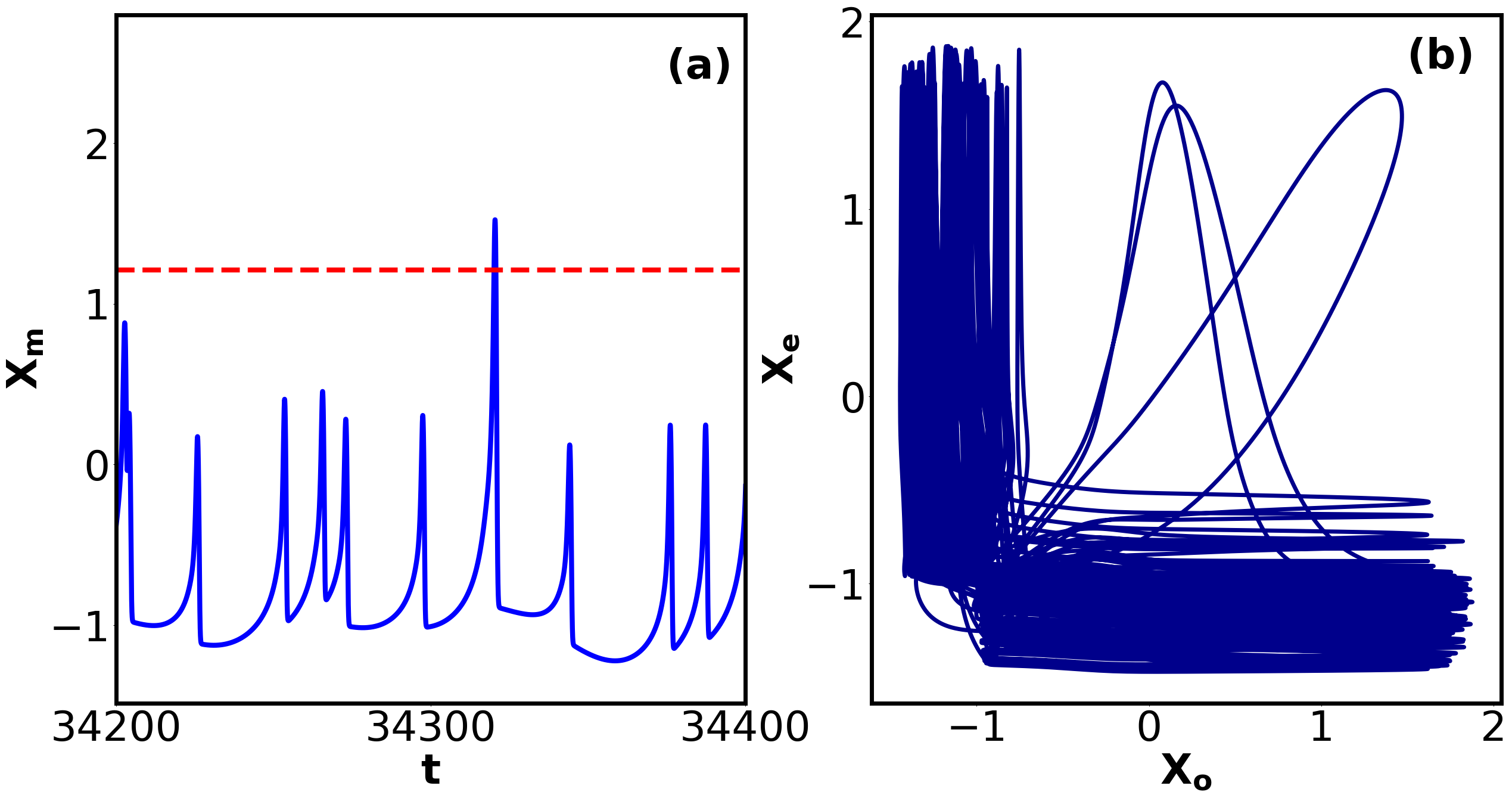}
		\includegraphics[width=1.05\linewidth]{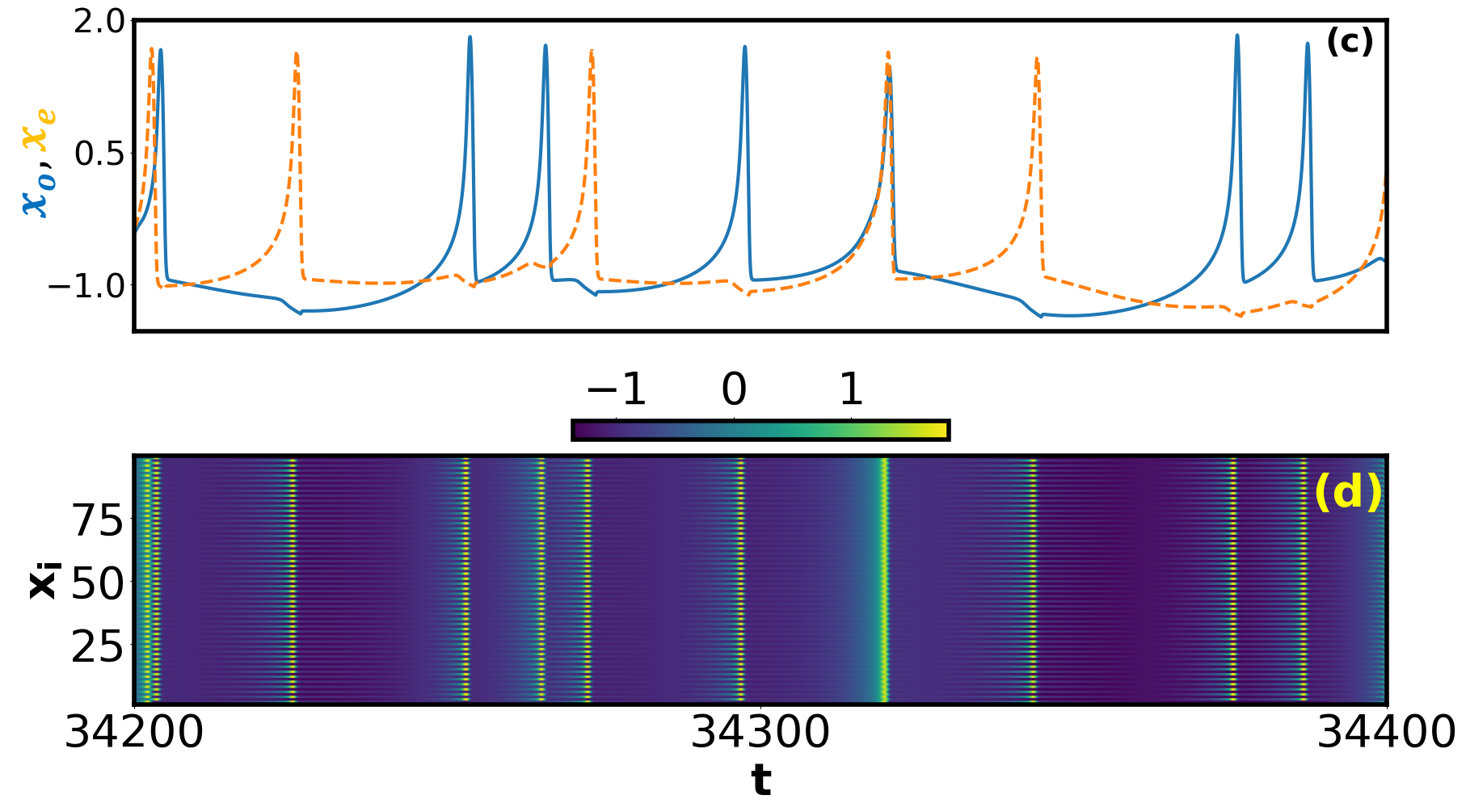}
		\caption{Macroscopic and microscopic dynamics of system (\ref{eqncoupled}) for $k=-0.17$. In particular, (a) $\&$ (b) are respectively the collective time series and synchronization error plot in the ($x_o-x_e$) plane. (c) represents the overlap of individual time series of $x_o$ (blue solid) and $x_e$ (red dashed). (d) is the spatio-temporal plot.}
		\label{eedy}
	\end{figure}
	
\begin{figure}
		\centering
		\includegraphics[width=1\linewidth]{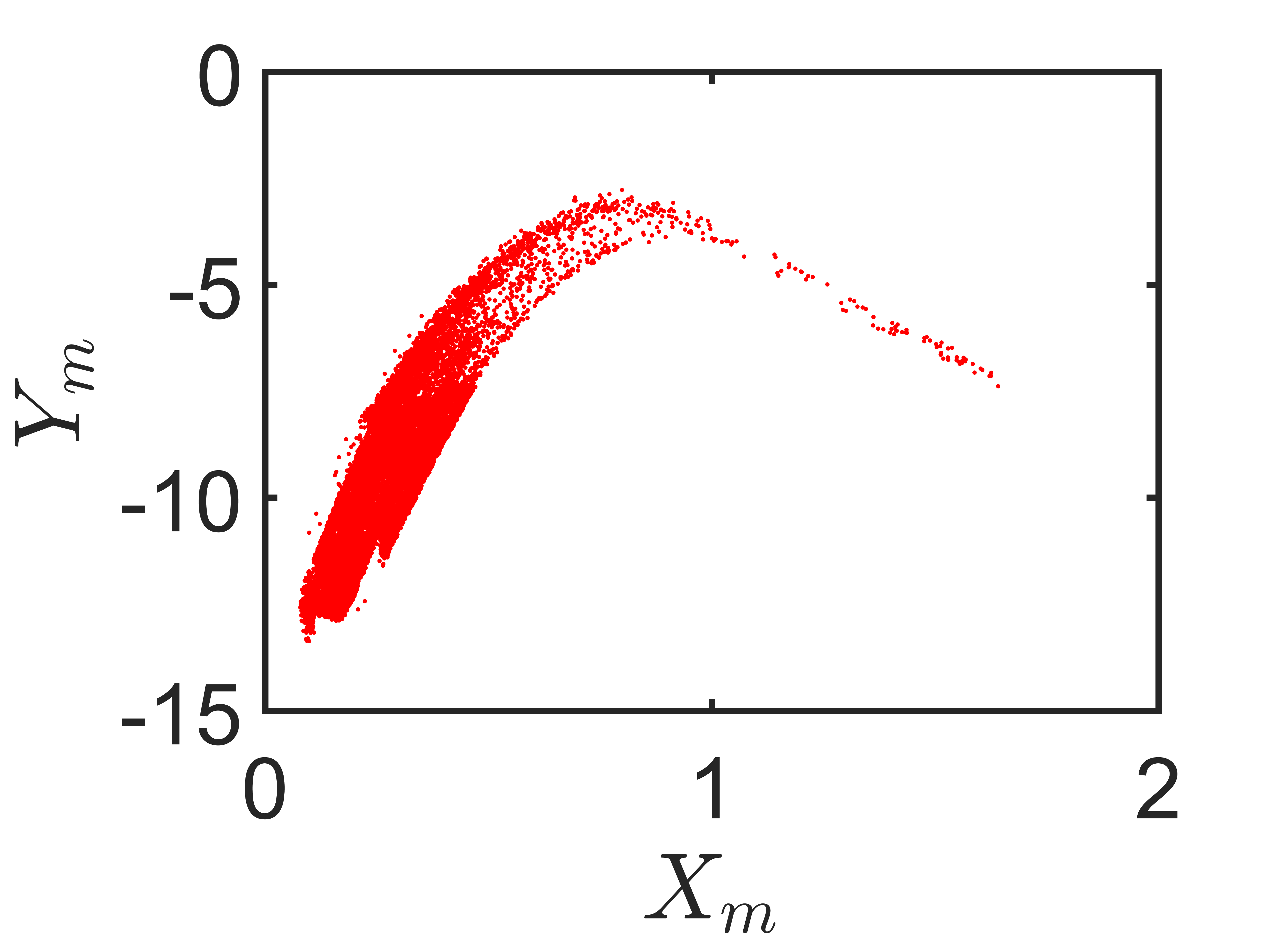}
		\caption{Poincar\'e cross section of the collective dynamics of system \ref{eqncoupled} greater than $X_m >0$ at $k=0.17$. This confims the chaotic nature of the system. The sparse points at the right end of the plot represents the rare extreme events.}
		\label{poincare}
	\end{figure}

	\textbf{(ii) $k=-0.17$:} Next, we move on to the region between $k=-0.23$ and $k=-0.06$ where we notice the evolution of EE in the collective observable $X_m$. To observe the dynamics of the system in this region, we choose $k=-0.17$. Similar to the previous case, here also we plot similar four different plots in Fig.~\ref{eedy} and analyze the dynamics. Upon inferring Fig.~\ref{eedy}(a), we can notice that the system exhibits chaotic spiking dynamics in the time series of $X_m$ and in the middle, large spikes rarely cross the threshold $H_s=m+6\sigma=1.23$. These rare occurrences of peaks crossing the $H_s$ (peak crossing the solid red-dashed line) correspond to EE. Here a chaotic spiking scenario occurs which confirms that within this zone of the coupling strength, $X_m$ is chaotic.  When we observe the dynamics of the two clusters ($x_o$ and $x_e$) in Fig.~\ref{eedy}(c), we can notice that both the clusters exhibit a chaotic bursting property and the number of spikes within bursts is irregular, and changes continuously from time to time. Importantly, they are out of phase with each other (blue solid line out of phase with orange dashed line in Fig.~\ref{eedy}(c)). But when we look at the time series of the individual clusters at the time of EE in $X_m$, to our surprise, we find that clusters phase synchronize with each other (in Figs.~\ref{eedy}(a) and \ref{eedy}(c)). Intermittent synchronization occurs between them, that is most of the time the clusters $x_o$ and $x_e$ are out of phase and only once in a while they phase synchronize with each other. Interestingly, whenever intermittent phase synchronization occurs between $x_o$ and $x_e$, an EE is found to occur in the $X_m$ observable. Even for a long range of the time series, rare emergence of EE is visible. From this, we can conclude that the single neuron exhibits chaotic bursting and collective observable exhibit chaotic spiking. The value of the largest Lyapunov exponent for the collective dynamics at $k=-0.17$ is found to be $1.0413$ confirming the chaotic nature of the dynamics. The corresponding Poincar\'e cross section at $k=-0.17$ is presented in Fig.~\ref{poincare}. Data points only for $X_m>0$ are plotted. The sparse points on the right side corresponds to the extreme events.
	
	Figure \ref{eedy}(b) shows us the projection of the antiphase synchronization manifold of $x_o$ and $x_e$ neurons. From the phase plane, we can observe that the clusters are in antiphase state most of the time, but occasionally they are in synchronized state and EE emerge. In this phase plane, the trajectories in the transverse direction are very sparse confirming the rarity in the phase synchronization. To validate the phase synchronization among all the neurons, during the emergence of EE, we plot the spatio-temporal plot in Fig.~\ref{eedy}(d). During all the time, odd and even clusters are not synchronized. But, at the time of the EE (peak crossing red dashed $H_s$ line in Fig.~\ref{eedy}(a)), correspondingly in the spatio-temporal plot (Fig. \ref{eedy}(d)),  we can observe that all the neurons are phase synchronizing with each other, where violet regions and green dots correspond to the resting and bursting states respectively. At the times of EE, we can notice a strong green line showcasing the synchronization among the neurons. It is also important to note that in the inhibitory coupling, the cluster synchronization is a phase synchronization, whereas when the coupling is excitatory, the cluster synchronization is complete synchronization.
	
	\begin{figure}
		\centering
		\includegraphics[width=1\linewidth]{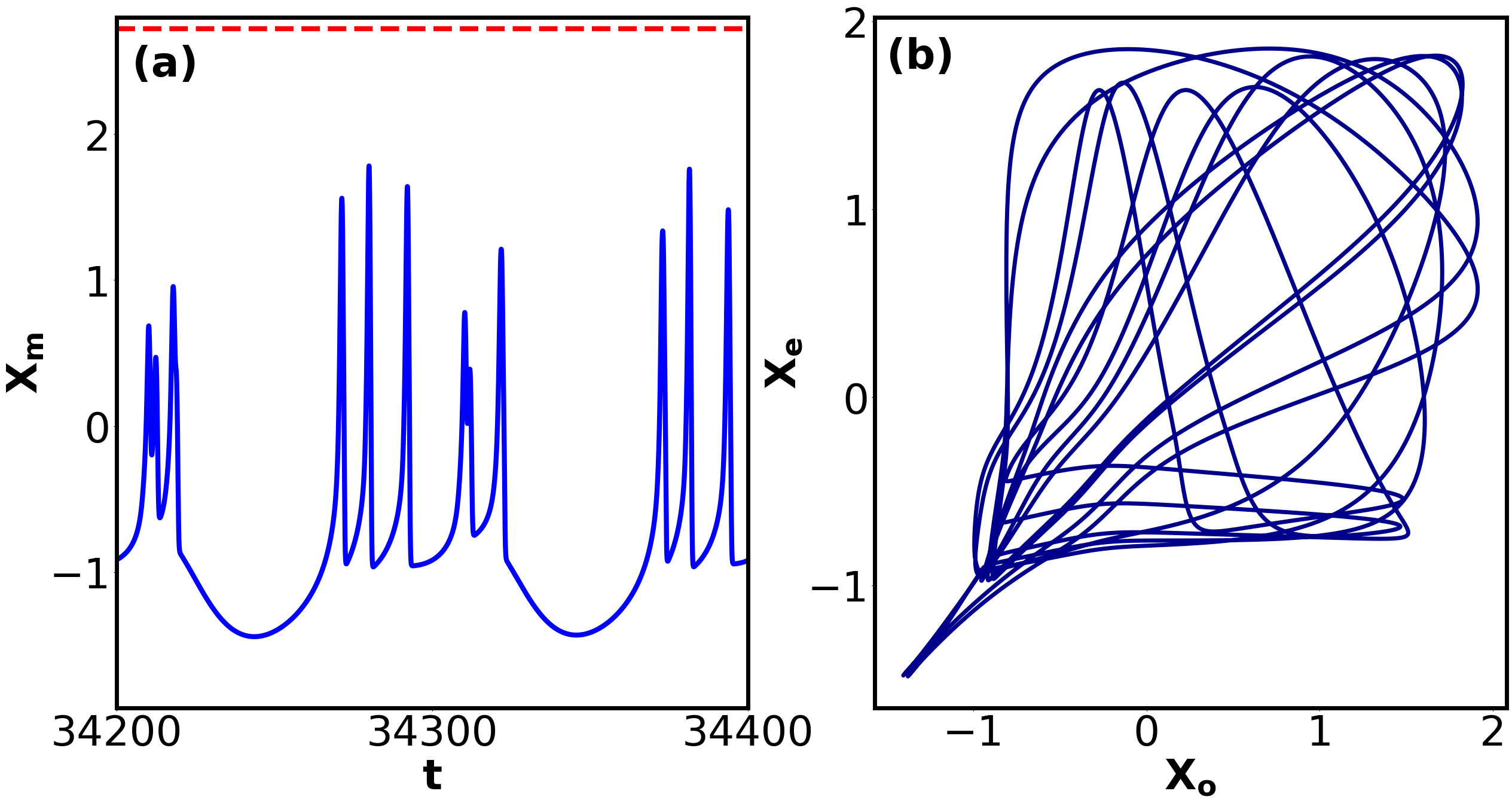}		\includegraphics[width=1.05\linewidth]{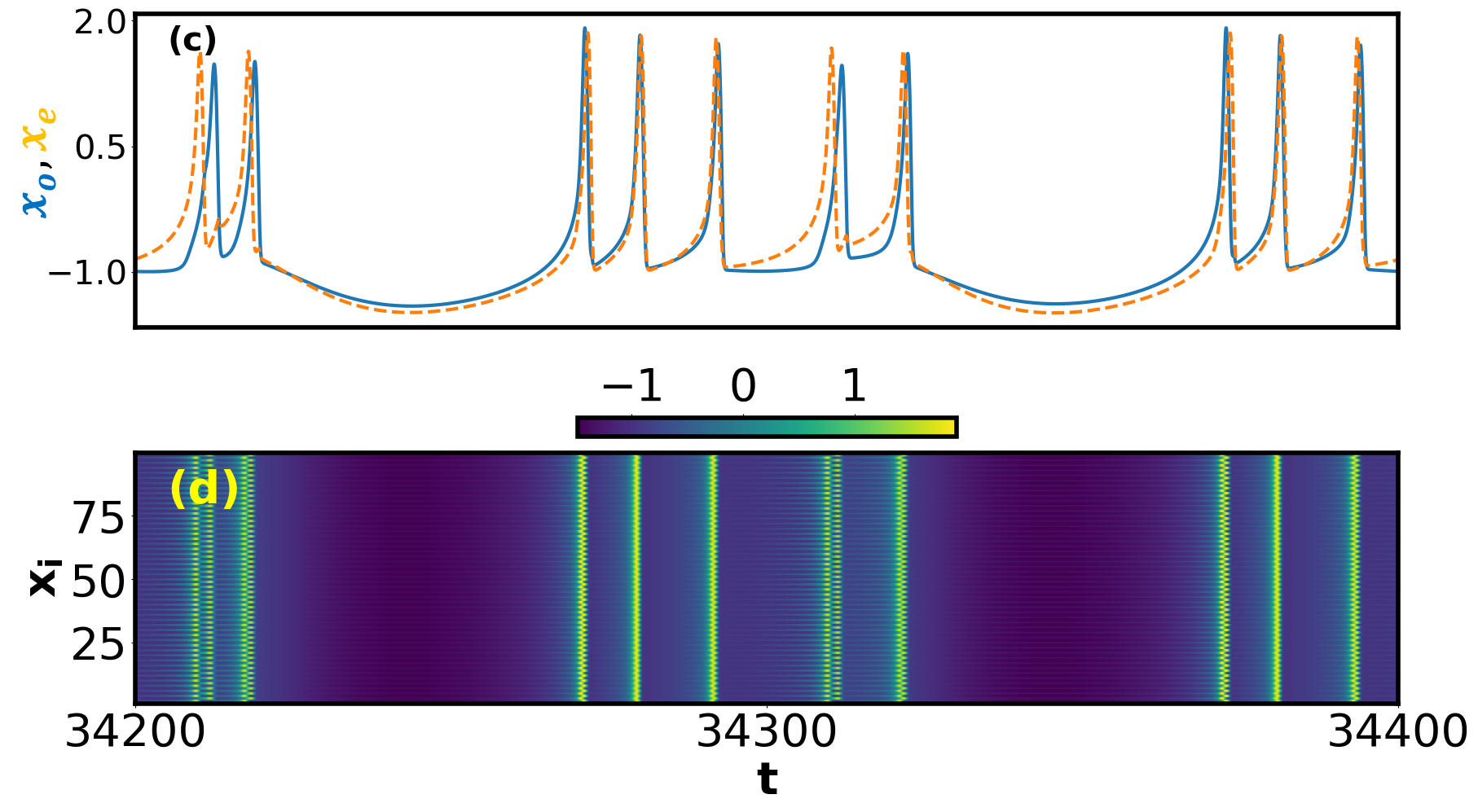}
		\caption{Macroscopic and microscopic dynamics of system (\ref{eqncoupled}) for $k=0.4$. In particular, (a) $\&$ (b) are respectively the collective time series and synchronization error plot in the ($x_o-x_e$) plane. (c) represents the overlap of individual time series of $x_o$ (blue solid) and $x_e$ (red dashed). (d) is the spatio-temporal plot.}
		\label{neesy}
	\end{figure}
	
	\par \textbf{(iii) $k=0.4$:} Next we choose another non-extreme region between $k=-0.5$ to $k=0.5$, namely, $k=0.4$ and try to understand the dynamics. So we plot in Fig.~\ref{neesy}(a), the time series of the collective observable $X_m$. We can observe that the collective dynamics of neurons shift from chaotic spiking to bursting property with five within-burst spiking. No peaks cross the threshold and hence no EE. The collective observable $X_m$ exhibits bursting. The time series of the individual clusters in Fig.~\ref{neesy}(c), shows the synchronization between $x_o$ and $x_e$ neurons. The amplitude may not match but they are in complete phase synchrony. No EE is evident in this region because as all the neurons get synchronized, the threshold value raises to a very high level. The synchronization of the odd and even neurons can also be observed from Fig.~\ref{neesy}(b). Since at all times, the trajectories are well synchronized, it moves only in the transverse direction. Finally, Fig.~\ref{neesy}(d) shows pretty much clearly how all the neurons get synchronized in most of the regions which clearly explains why we are not getting EEs in Fig.~\ref{neesy}(a). Here we can spot a bright green lines throughout the time which confirms the synchronization. 

    \par Analysis done so far, reveals that for a unidirectional ring of HR neurons with clustered ICs, varying the coupling strength transits the oscillators present in the system from anti-phase to in-phase synchronization. This transition happens gradually with the occurrence of occasional in-phase synchronization in time. As the coupling strength is further increased, occasional in-phase synchronization becomes frequent  and for higher values of the coupling strength, burst synchronize occurs between them. Thus on the route to synchronization, there are regions of the coupling strength where intermittent synchronization occur among the clusters. During such an intermittent phase synchronization, EE emerge in the collective observable. This intermittent phase synchronization between two cluster of oscillators (intermittent cluster synchronization) is the potential reason behind the emergence of EE in the considered system. It is well-known that systems with symmetries will undergo a transition from antiphase to in-phase synchronized state upon varying the coupling strength \cite{gb1,gb2,gb3,gb4,gb5}. But in the present scenario, independent of the number of clusters of ICs, intermittent phase synchronization between the clusters occur leading to EE. Two things are important to be noted here. (i) All these results were obtained for $10^9$ iterations after leaving out $3\times10^5$ iterations as transients. (ii) This phenomena is true even when the ICs was randomly picked from the phase portrait basin. We checked the dynamics for 100 realizations and for all, qualitatively the dynamics was found to be the same. These two properties confirm that the intermittent cluster synchronization phenomenon that take place in the system is a well settled structurally stable behavior and not a transient. Further, this phenomenon is found to be robust with respect to the network size. Further, it is to be noted that similar phenomenon is not observed for non-local or global network topology. In these cases, chaotic dynamics existed in the mean field observable, but no EE are observed.

	\section{Robustness of Cluster synchronization}
	Due to the choice of ICs, intermittent two cluster synchronizations are found to happen in the system (\ref{eqncoupled}) and as an effect, the system collectively is found to exhibit EE. In other words, we used two clusters of initial conditions due to which system (\ref{eqncoupled}) evolves as a two-body problem that is odd neurons and even neurons got synchronized separately. Intermittent synchronization of these clusters leads to the emergence of EE in the chaotic region. Now let us discuss about the robustness of this cluster synchronization with respect to the increase in the number of clusters of initial conditions. We have found that when the number of clusters of initial conditions given to the system increases, correspondingly the number of clusters that the system splits into also increases. To prove this, we demonstrate below the phase synchronization of clusters using the spatio-temporal plots.
	\par
\begin{figure}[h!]
		\centering
		\includegraphics[width=1\linewidth]{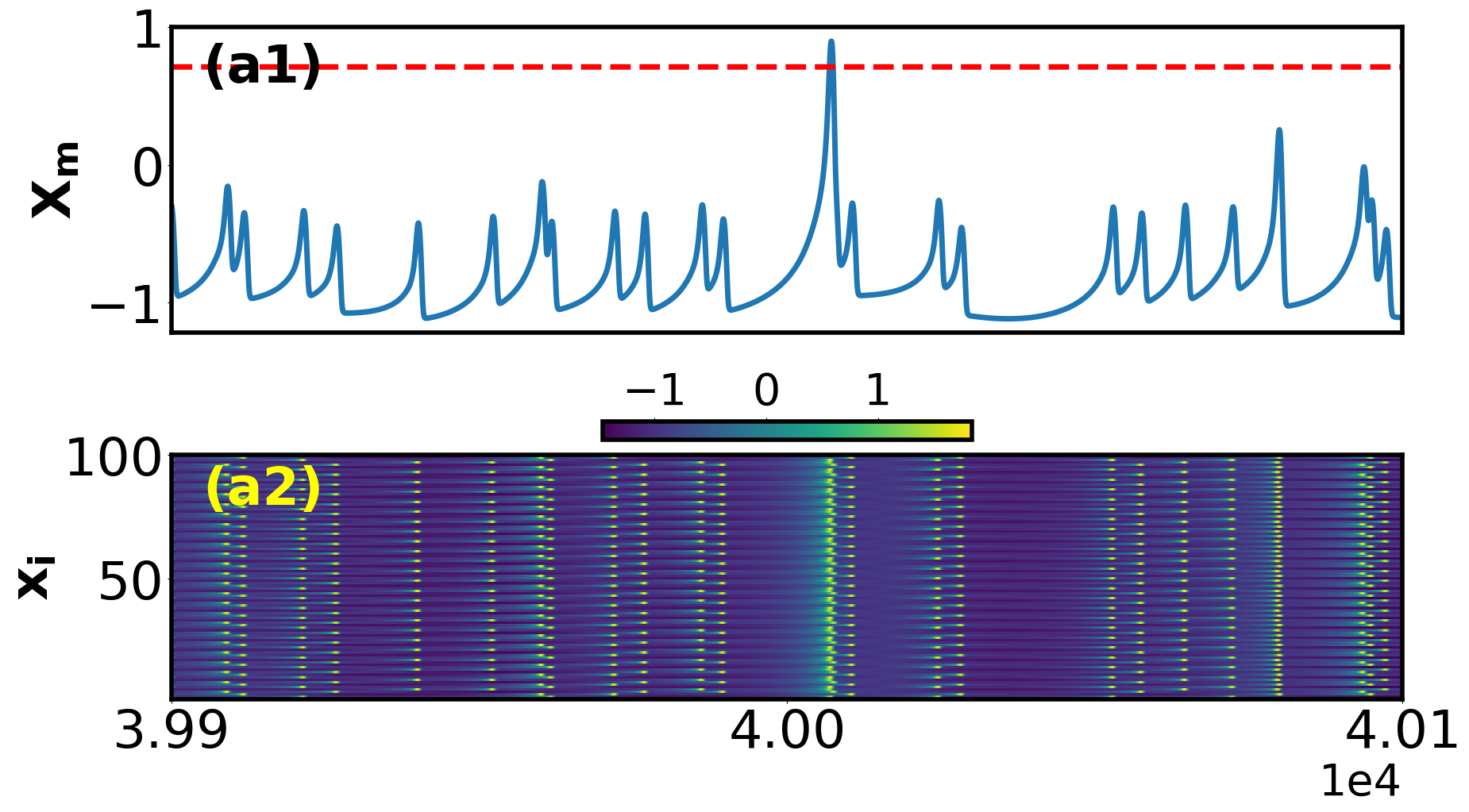}
		\includegraphics[width=1\linewidth]{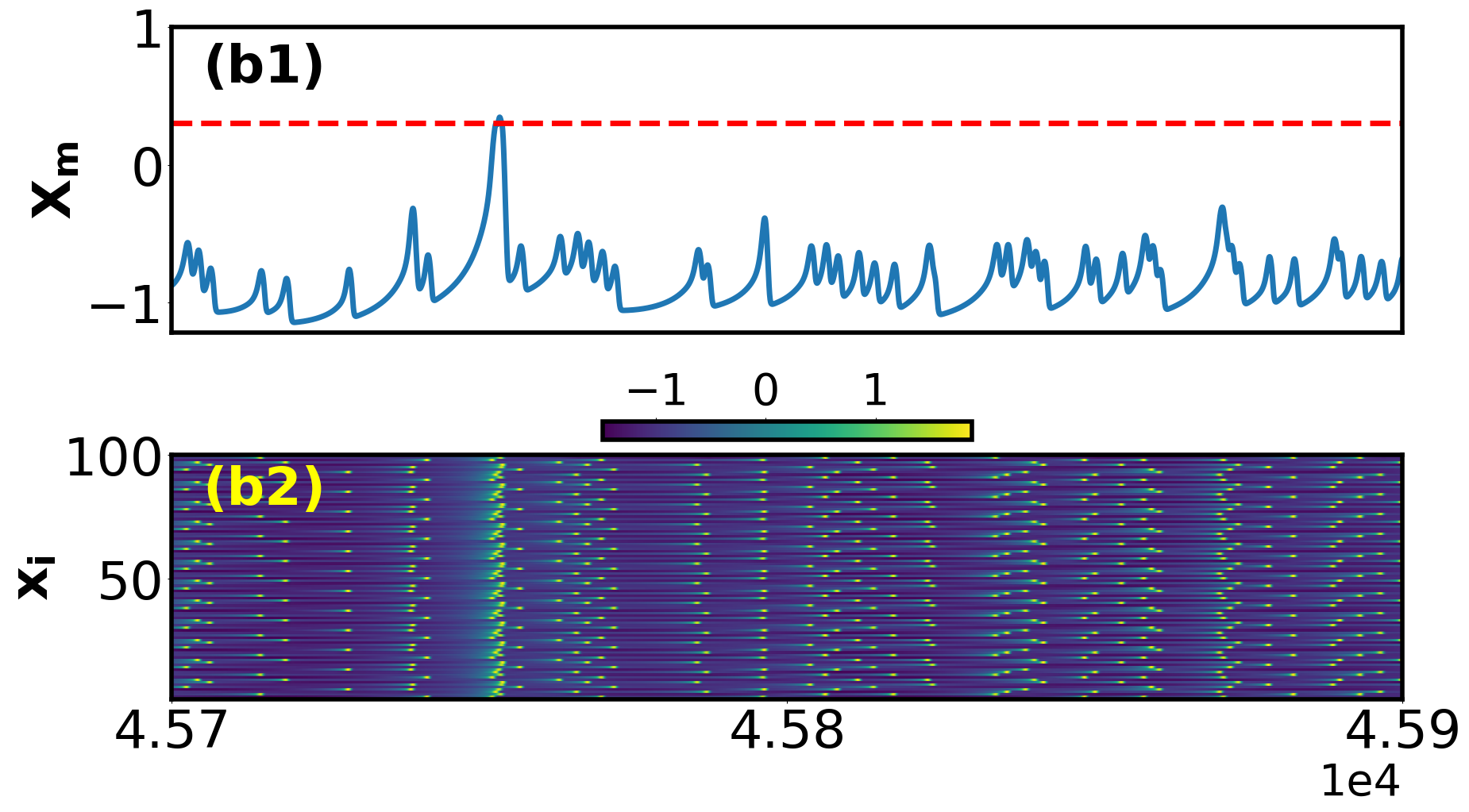}
		\includegraphics[width=1\linewidth]{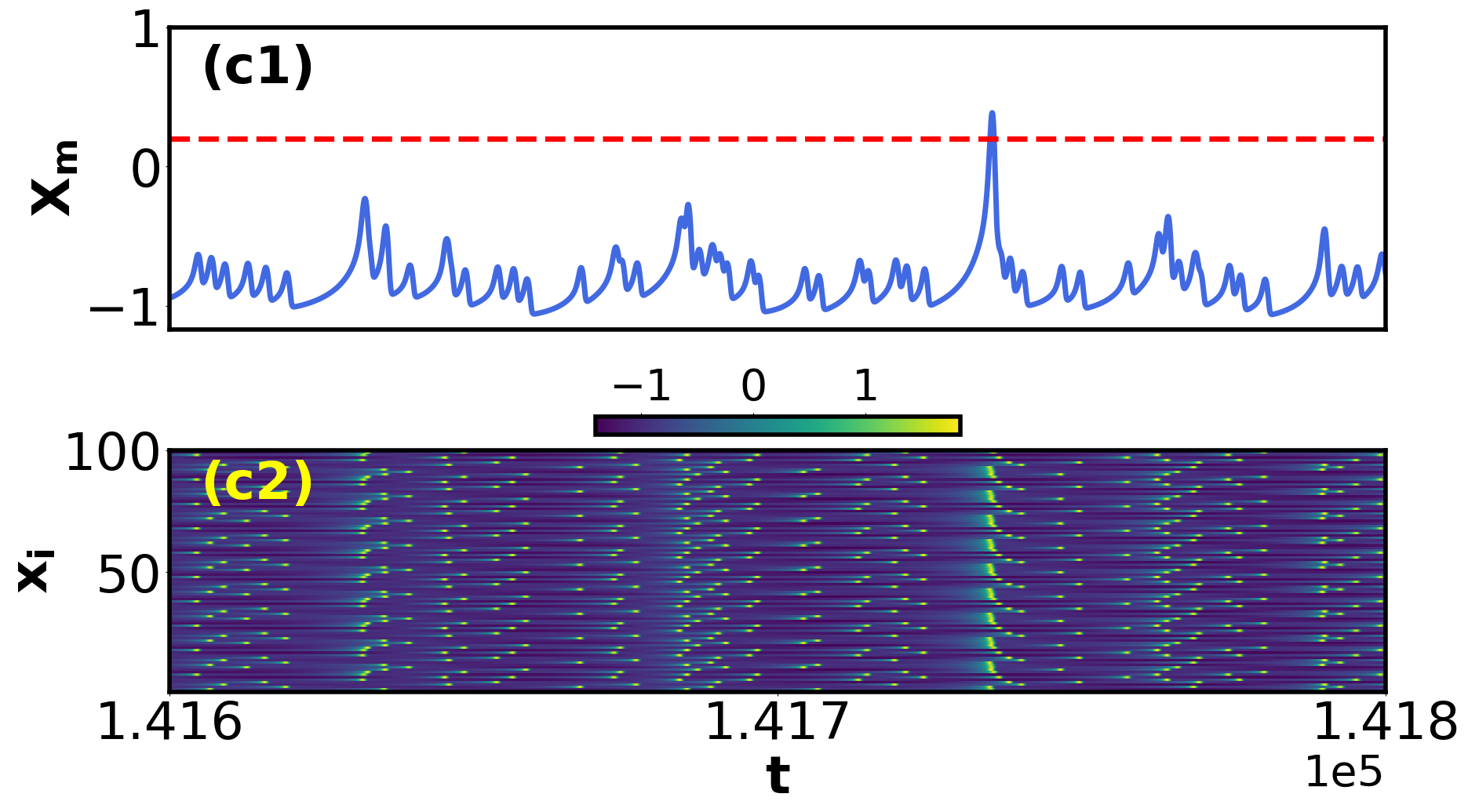}
		\caption{(a1,a2), (b1,b2) and (c1,c2) shows the dynamics of system (\ref{eqncoupled}) at $k=-0.17$ respectively for 4, 8 and 10 different ICs. Subplots with index 1 are the time series and subplots with index 2 are spatio-temporal plots.}
		\label{fig:20b}
	\end{figure}

 Figures~\ref{fig:20b}(a), \ref{fig:20b}(b) and \ref{fig:20b}(c) show the dynamics of system (\ref{eqncoupled}) respectively for 4, 8 and 10 ICs. Subplots with index 1 $(a1,b1,c1)$ are the time series and subplots with index 2 $(a2,b2,c2)$ are spatio-temporal plots. First for the 4 ICs case, by observing Figs.~\ref{fig:20b}(a1) and \ref{fig:20b}(a2) at the EE occurring part, three out of four clusters get involved in synchronization without the involvement. Next for 8 ICs, in Figs.~\ref{fig:20b}(b1) and \ref{fig:20b}(b2)  at the EE region, only five clusters get synchronized whereas three clusters are not synchronized. Finally for the 10 ICs case, in Figs.~\ref{fig:20b}(c1) and \ref{fig:20b}(c2), at the EE point, we can observe that only seven clusters get synchronized while three clusters do not get synchronized. In addition, we have also checked for 100 different ICs from random generation. We find that in this case, if 25-30 oscillators come in a very small time zone, EE occur in the collective variable.
	
\begin{figure}[!ht]
		\centering
		\includegraphics[width=1.0\linewidth]{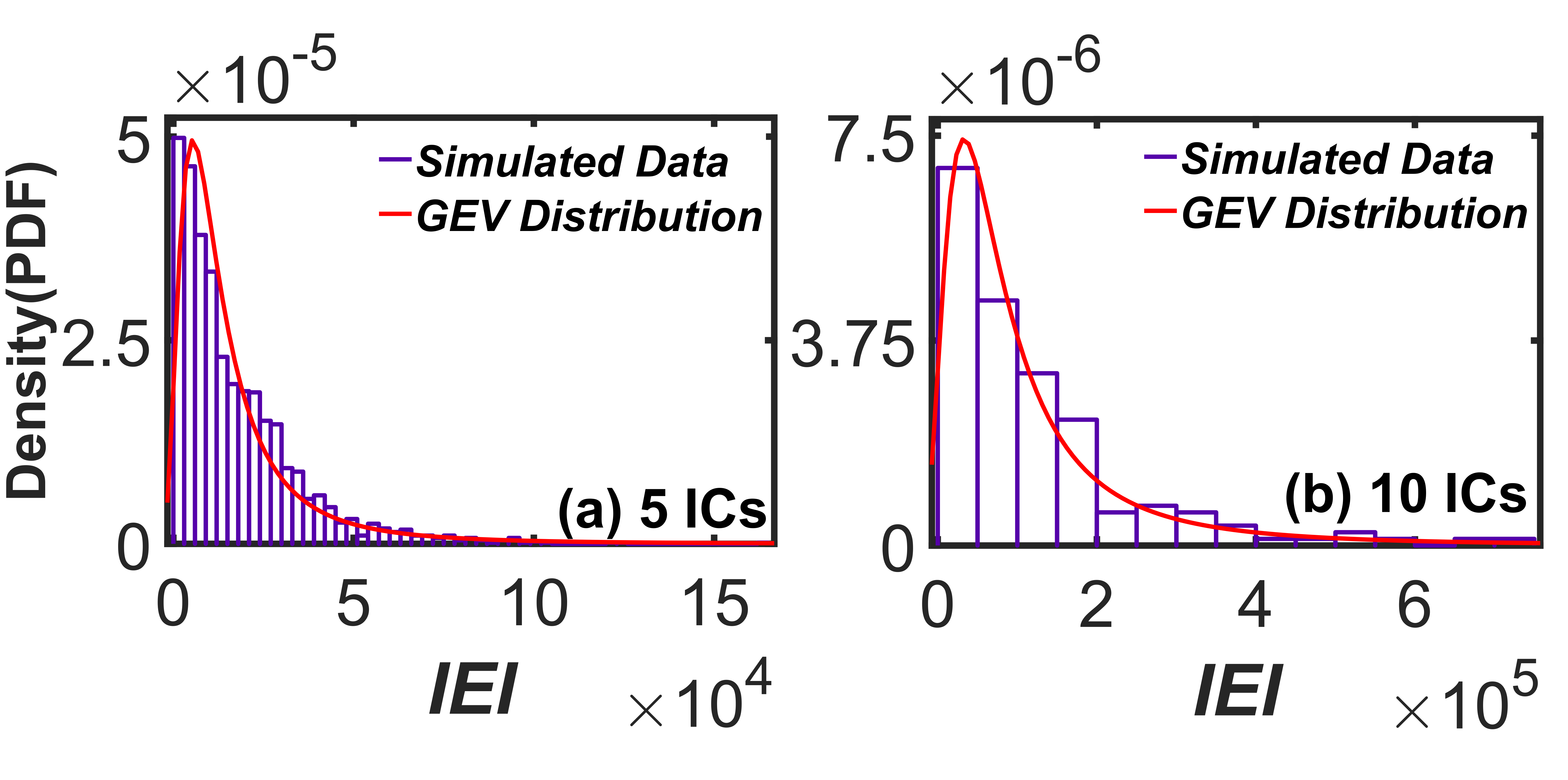}
		\caption{IEI distribution of system (\ref{eqncoupled}) at $k=-0.17$ for (a) 5 and (b) 10 different ICs. Blue line is the histogram of the simulated data whereas red line is the fitted GEV distribution. }
		\label{icd}
	\end{figure}

\begin{figure}[!ht]
		\centering
		\includegraphics[width=1.0\linewidth]{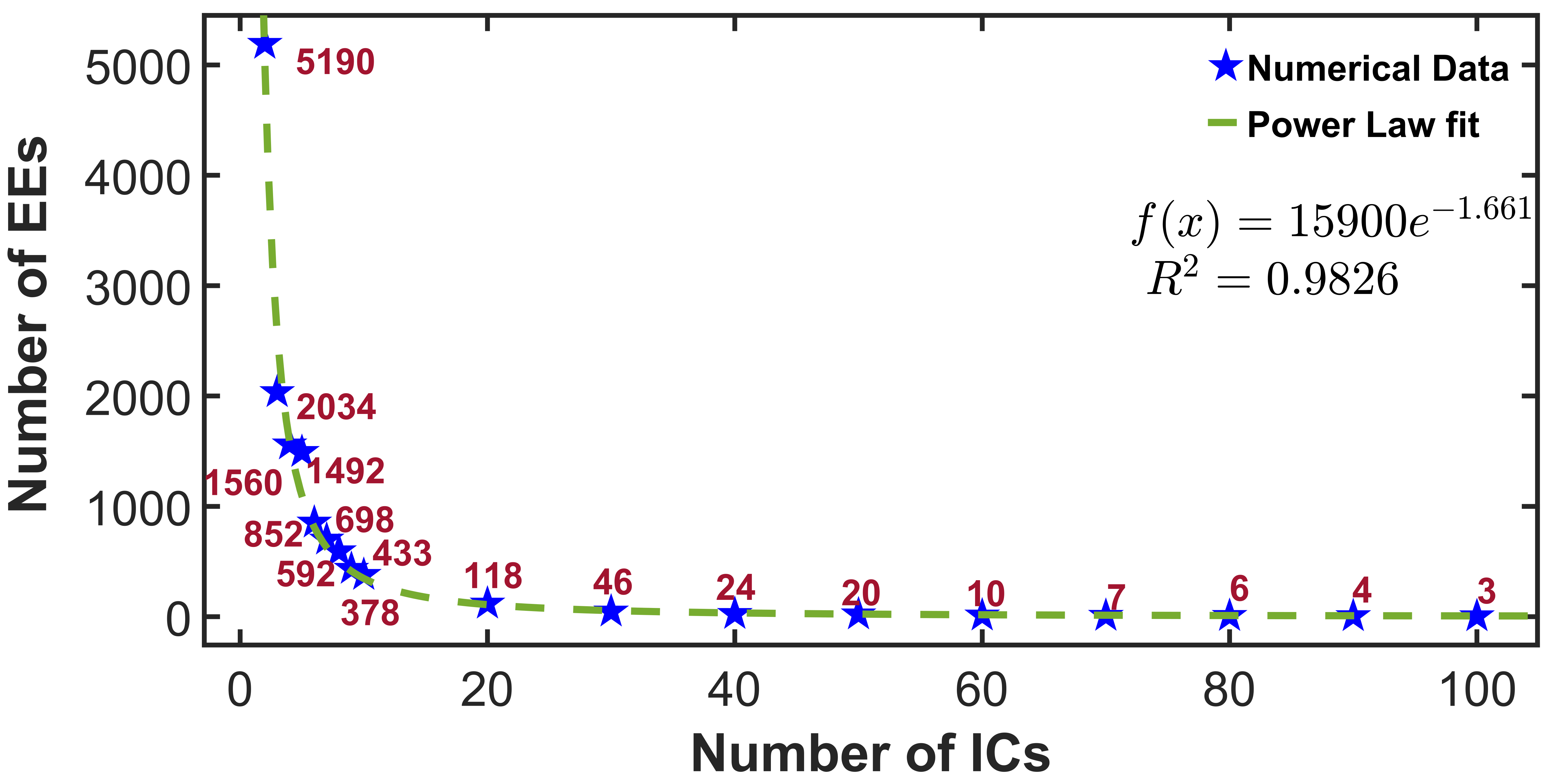}
		\caption{Number of EE peaks versus the number of different ICs in system (\ref{eqncoupled}) at $k=-0.17$. Here legend blue star represents the number of EE calculated numerically, the green dashed curve denotes the power law fit.}
		\label{ics}
	\end{figure}

	As the number of clusters increases the occurrence of EE will become even more rarer. This is because intermittent synchronization between the clusters will become rarer and rarer. This can be inferred from the IEI plots in Fig.~\ref{icd} where the IEI distribution is fitted with generalized extreme value distribution (GEV). Initially for two ICs, we can notice the maximum time interval between the two EE will be 30,000 (see Fig.~\ref{fig:12}(b)). Next for 5 ICs, from Fig~\ref{icd}(a), we can observe that the maximum time interval is $16\times 10^4$. Figure~\ref{icd}(b) is the IEI for 10 different ICs where the maximum time interval is $70\times 10^4$. This shows that the interval between two EE increases as the number of different ICs increases.  Similarly, for 100 ICs case, in the entire $10^9$ iterations, only 3 EE were found to occur. To explain this, we plot in Fig.~\ref{ics}, the number of EE peaks versus the number of ICs. In Fig.~\ref{ics}, $x$ and $y$ axes denote the number of ICs and number of EE respectively. Legend blue star represents the number of EE calculated numerically while the green dashed curve denotes the power law fit. The expression for the fit and the corresponding coefficient of determination $R^2$ values are given in the figure itself. From the plot, we can infer decrease in EE with an increase in ICs. That is ICs and EE are inversely proportional. This confirms the raise in the rarity of the EE with the number of number of ICs.
	
	\section{Conclusion}
	The collective dynamics of the membrane potential of a network consisting of $N$ unidirectionally coupled HR neurons, for a range of coupling strength, is found to spike occasionally to higher amplitudes that is $6$ times its standard deviation from the mean level. These rare spikes display dragon king-like statistics and a long tail in the non-Gaussian distribution substantiating that these are EE. Observing the two-parameter phase diagram between the coupling strength and input current, reveals the regions that exhibit EE. During the increase in the coupling strength, we also found that the collective observable transits from complete chaotic spiking to mixture of chaotic spiking and bursting to complete bursting. During this transition, EE occur when the collective dynamics is either a chaotic spiking or a mixture of chaotic spiking and bursting. During such a mixed dynamics, the EE occur as a result of intermittent cluster synchronization at the microscopic level. Due to the choice of ICs, odd neurons and even neurons tend to form separate clusters. These clusters are mostly out of phase with each other and only intermittently they are in phase synchronized state. During such intermittent phase synchronization, EE occur in the collective observable. Apart from this, increasing the number of different ICs increases the number of clusters in the system. For 100 different ICs, it can be observed that if 25-30 neurons get into a very short time zone, EE are produced. IEI plots and number of EE plot deciphers the rarity of the events for an increase in ICs. 
	
	\par Despite several mechanisms to EE are explored for lower dimensional networks, very little has been explored about the route to EE in higher dimensional coupled systems. Especially, no routes to EE in higher dimensional HR network has been determined so far. In this work, we have shown a new mechanism called the intermittent cluster synchronization as a new route of EE in unidirectionally and locally coupled HR neurons. Intermittent cluster synchronization biologically refers to the sudden synchronization occurring among the neurons in the brain, collectively leading to a detrimental disorder like, epilepsy. Possible mitigation strategies like occasional random perturbations, pinning of oscillators, introducing other specific forms of coupling may serve as a new avenue in preventing this detrimental and unwanted event. Some possible  Importantly, present result and the mechanism found out in this work serves as a milestone in the understanding of dynamical mechanisms behind epileptic seizures. 
    \par In this work, we observed EE in a symmetric network and intermittent cluster synchronization is observed irrespective on the symmetric in the initial conditions. In this connection, the symmetry breaking bifurcations in equivariant systems \cite{gb1,gb2,gb3} could be an interesting problem in future to (i) identify the subgroup symmetries or equivariant bifurcations (e.g., Hopf or pitchfork) associated with the cluster states, and (ii) explore the emergence of anti-phase and in-phase synchronization as solutions constrained by the Cn symmetry group of the ring topology. 
	
	\section*{Acknowledgements}
	S.S. thanks the Science and Engineering Research Board (SERB), Department of Science and Technology (DST), Government of India for providing financial support in the form of National Post-Doctoral Fellowship (File No.~PDF/2022/001760).  The work of M.S. forms a part of a research project
	sponsored by the Council of Scientific and Industrial Research (CSIR) under Grant No. 03/1482/2023/EMR-II. D.G. is supported by DST-SERB Core Research Grant (Project No. CRG/2021/005894). S.S. thanks Tapas Kumar Pal and Dhiman Das for their benevolent discussions and valuable feedback on this manuscript in particular on the statistical part. 
	
	\section*{Appendix A} 
\begin{figure}[!ht]
		\centering
		\includegraphics[width=1\linewidth]{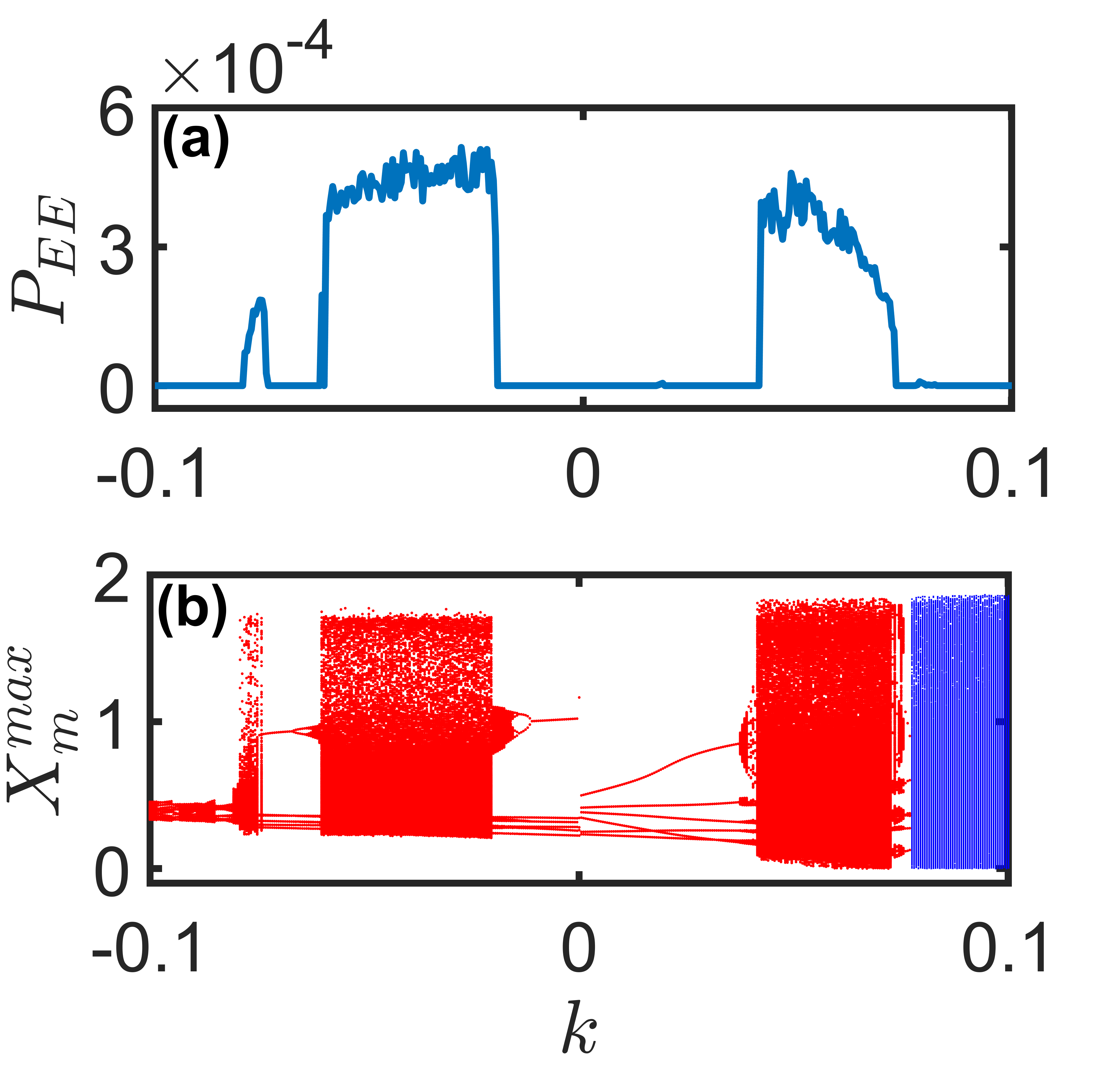}
		\caption{(a) The values of probability $P_{EE}$ and (b) bifurcation diagram of $X_m$ of system (\ref{eqncoupled1}) by varying coupling strength $k$ at $I=4$. Blue points in (b) represents chaotic regime with out 
 extreme events. Here the periodic  and chaotic regimes (red points) show non-EE and EE due to anti-phase and occasional in-phase synchronization between two clusters.}
		\label{fig:22}
	\end{figure}

	Another kind of interaction commonly used to model the nearest neighbour interaction in network theory of coupled oscillators is the diffusive coupling. We analyze the emergence of EE in a unidirectional ring of diffusively coupled HR neurons. The corresponding equation of motion is
	\begin{equation}
		\begin{split}
			\dot{x}_i&=y_i+bx_i^2-ax_i^3-z_i+I-k(x_{i+1}-x_i), \\
			\dot{y}_i&=c-dx_i^2-y_i,\\
			\dot{z}_i&=r[s(x_i-x_r)-z_i],\\
			\label{eqncoupled1}
		\end{split}  
	\end{equation} 
	where $i=1,~2,~\hdots,~N$ is the number of neurons in the system with the boundary condition ($x_{N+1}=x_1$). The definition and values of the parameters are the same as in Sec.~2. Similar to the chemical coupling case, in the diffusive coupling case also we observe the EE to occur in the system (\ref{eqncoupled1}) for various values of the coupling strength $k$. The probability of EE and bifurcation diagram with respect to coupling strength $k$ are shown in  Figs.~ \ref{fig:22}(a) and Fig.~\ref{fig:22}(b), respectively. Comparing both these figures, we can see that the whenever the system exhibits non-zero probability (Fig.~\ref{fig:22}(a)), correspondingly the size of the attractor is also high. Wherever the system is periodic, anti-phase synchronization occurs, resulting in the non-observance of extreme events. This happens throughout the parameter space $k$, except for higher values of the coupling strength (shown by blue dotted points in Fig.~ \ref{fig:22}(b)) where burst synchronization occurs, and no extreme events appear in the system.
	
	\begin{figure}[!ht]
		\centering
		\includegraphics[width=1\linewidth]{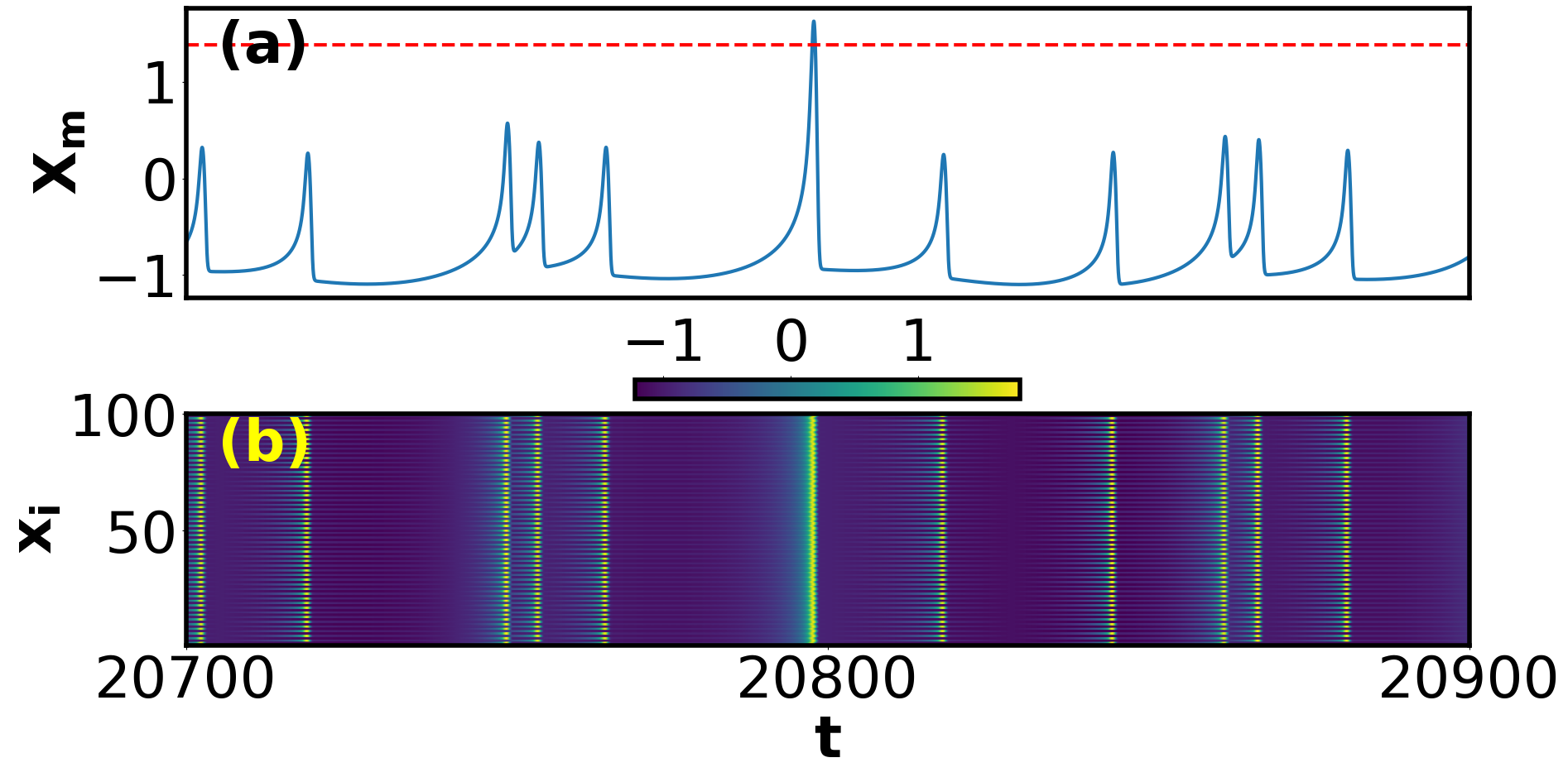}
		\caption{(a) Time series and (b) spatio-temporal plot of system (\ref{eqncoupled1}) at $k=-0.018$.}
		\label{fig:28}
	\end{figure}
	
	In the repulsive coupling, EE are found in the region between $k=-0.079$ and $k=-0.015$. Also for the attractive coupling, EE are found between $k=0.014$ and $k=0.072$. To visualize, we plot the time series of the collective observable for $k=-0.018$ in Fig.~\ref{fig:28}(a)  where the trajectories cross the threshold $H_s=1.38$. The mechanism through which EE occur in the system (\ref{eqncoupled1}) is also intermittent cluster synchronization. To demonstrate this, in Fig.~\ref{fig:28}(b), we plot the spatio-temporal plot. Cluster synchronization can be seen by the bright green line in the spatio-temporal plot corresponding to the EE in the collective observable $X_m$.
	
		\begin{figure}[!ht]
		\centering
		\includegraphics[width=1\linewidth]{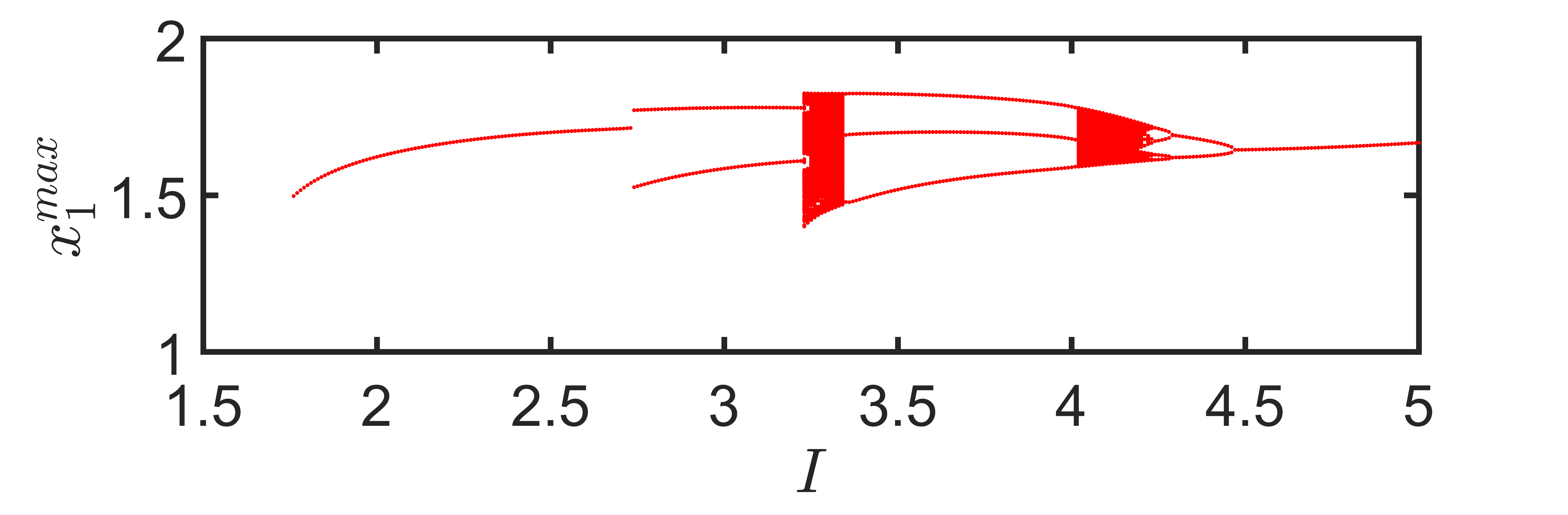}
		\caption{Bifurcation diagram of node-1 of system (\ref{eqncoupled1}) with respect to $I$. Here period-doubling route to chaos is observed by decreasing the value of $I$.}
		\label{fibifurdiff}
	\end{figure}

In Fig.~\ref{fibifurdiff}, we show the dynamics of node 1 in the form of bifurcation diagram. We can observe that periodic nature interspersed by chaotic behaviour. This chaotic behaviour is observed by period doubling route.

\end{document}